\documentclass[11pt]{article}

\usepackage{cite}
\usepackage{color}
\usepackage{amsfonts}
\usepackage{amsmath}
\usepackage{epsfig}
\usepackage{latexsym}
\usepackage{fancybox}
\usepackage{graphicx}
\usepackage{subfigure}
\usepackage{amssymb}
\usepackage{ascmac}
\numberwithin{equation}{section}
\allowdisplaybreaks

\def\spa#1{\phantom{\fbox{\rule[-#1cm]{0cm}{0cm}}}}
\def\bal#1\eal{\begin{align}{#1}\end{align}}
\def\bitem#1\eitem{\begin{itemize}{#1}\end{itemize}}
\def\benu#1\eenu{\begin{enumerate}{#1}\end{enumerate}}

\newcommand{\hs}{\hspace}
\newcommand{\vs}{\vspace}
\newcommand{\del}{\partial}
\newcommand{\nn}{\nonumber}

\newcommand{\mc}{\mathcal}

\newcommand{\what}{\widehat}

\newcommand{\ol}{\overline}
\newcommand{\wtilde}{\widetilde}
\newcommand{\mbb}{\mathbb}
\newcommand{\ul}{\underline}

\def\[#1\]{\begin{align}#1\end{align}}

\begin{document}

\hfuzz=100pt
\title{{\Large \bf{ 
Localization of Gauged Linear Sigma Model
for KK5-branes
}}}
\author{Yuki Hiraga$^{a}$\footnote{hiraga@eken.phys.nagoya-u.ac.jp} and Yuki Sato$^{a,b}$\footnote{ysato@th.phys.nagoya-u.ac.jp }
  \spa{0.5} \\
\\
$^a${\small{\it Department of Physics, Nagoya University}}
\\ {\small{\it Chikusaku, Nagoya 464-8602, Japan}}\\
$^b${\small{\it Institute for Advanced Research, Nagoya University}}
\\ {\small{\it Chikusaku, Nagoya 464-8602, Japan}}
\spa{0.5} 
}
\date{}

\maketitle
\centerline{}

\begin{abstract}  
We study quantum aspects of the target space of the non-linear sigma model which is a low-energy effective theory of the gauged linear sigma model (GLSM). 
As such,  
we especially compute the exact sphere partition function of GLSM for KK$5$-branes whose background geometry is a Taub-NUT space, 
using the supersymmetric localization technique on the Coulomb branch.
 
From the sphere partition function, 
we distill the world-sheet instanton effects.     
In particular, we show that concerning the single-centered Taub-NUT space, the instanton contributions exist only if the asymptotic radius of the $S^1$ fiber in the Taub-NUT space is zero. 
\end{abstract}

\renewcommand{\thefootnote}{\arabic{footnote}}
\setcounter{footnote}{0}

\newpage

\section{Introduction}
\label{sec:introduction}
 
In string theory, lots of spatially extended objects, dubbed as branes, are included as dynamical degrees of freedom.   
Amongst all these branes, D-branes are the standard branes while there exist exotic branes characterized by co-dimensions and the brane tension; 
they are intimately related each other through various duality transformations \cite{deBoer:2012ma, deBoer:2010ud, Obers:1998fb}. 
A well-known example is a smeared NS$5$-brane (an H-monopole) that is T-dual to a KK$5$-brane (a Kaluza-Klein monopole). 
This relation can be directly checked, focusing on a harmonic function of the brane solution in supergravity and implementing the T-duality \cite{Buscher:1987sk} along the $S^1$-direction that takes on isometry in the background geometry of the H-monopole \cite{Kimura:2014wga, Tong:2002rq, Harvey:2005ab, Kimura:2013zva}. 
The background geometry of the KK$5$-brane obtained by the T-dualized H-monopole in this way is called a Taub-NUT space. 
This $4$-dimensional space is locally described as $\mbb{R}^3 \times S^1$ and is the $S^1$-fibration over the base space, $\mbb{R}^3$. 
In addition, one can obtain the exotic $5^2_2$-brane spacetime if, in $\mbb{R}^3$ of the KK$5$-brane, one performs the same procedure as in the case of the KK$5$-brane obtained through the H-monopole. 
This space is the $T^2$-fibration of the base space, $\mbb{R}^2$ \cite{deBoer:2012ma, deBoer:2010ud, Kikuchi:2012za, Kimura:2013zva}. 
The discussion above is based on supergravity theories that are low-energy effective theories of string theory, 
i.e. based on a point-particle picture in the limit that decouples massive modes in the string spectrum. 

The background geometries discussed above are known to get corrections via world-sheet instantons that are stringy effects \cite{Tong:2002rq, Okuyama:2005gx, Harvey:2005ab, Kimura:2013fda, Kimura:2013khz, Kimura:2018ain}. 
In order to understand such stringy effects, one need to reframe the spacetime structure from the point of view of not the point-particle picture but the string-theory picture. 
It is gauged linear sigma models (GLSMs) that allow one to carry out analyses through a perspective of world-sheet sigma models. 
GLSMs are $2$-dimensional supersymmetric gauge theories that can be interpreted as the UV-theories of the world-sheet non-linear sigma models (NLSMs). 
GLSM at UV is characterized by the moduli space of the supersymmetric vacuum that corresponds to the target space of NLSM at IR. 
It is remarkable that GLSM allows one to interpret geometric information in terms of the language of field theory. 
If a target space satisfies the Ricci-flatness condition, NLSM has conformal symmetry, 
meaning that GLSM has an aspect of superconformal field theory. 
Even in the context of GLSMs that are the UV theories, the T-dualities of NLSMs are realized in the superfield formalism \cite{Hori:2000kt}.  
Using the T-duality transformations in GLSMs, the $4$-dimensional target spaces of the H-monopole, the KK$5$-brane and the $5^2_2$-brane have been reformulated \cite{Tong:2002rq, Okuyama:2005gx, Harvey:2005ab, Kimura:2013fda, Kimura:2013khz, Kimura:2018ain}. 
The important point here is that brane solutions in supergravity theories based on the point-particle picture are indeed described in terms of the string-theory perspective.
Moduli that specify the location of centers in the target space are introduced as the FI- and $\theta$-parameters in GLSM \cite{Okuyama:2005gx}. 
Instantons characterized by winding numbers appear as configurations that satisfy a vortex equation in GLSM \cite{Witten:1993yc, Hori:2000kt, Tong:2002rq, Schroers:1996zy}.             
This instanton effect is nothing but the stringy effect that we are interested in.                       
 
From the sphere partition function of GLSM, one can read off information of the target space. 
The sphere partition function of GLSM is related to the K\"ahler potential on the moduli space of the target space, 
which was pointed out in \cite{Jockers:2012dk} and shown explicitly in \cite{Gomis:2012wy, Gerchkovitz:2014gta}. 
Superconformal field theories admit exactly marginal deformations, under which superconformal symmetries are preserved. 
In the context of the supersymmetric NLSM that is a low-energy effective theory of GLSM, this kind of modification of the theory means to keep the Ricci-flatness condition. 
Therefore, the exactly marginal deformation corresponds to the moduli space of the Ricci-flat manifold. 
For that reason, through the Zamolodchikov metric that is a metric on the conformal manifold \cite{Zamolodchikov:1986gt}, 
the sphere partition function of GLSM is closely tied to the K\"ahler potential \cite{Gomis:2012wy, Gerchkovitz:2014gta, Doroud:2013pka}. 

In addition, there exists an advantage to studying the sphere partition function of GLSM, i.e. 
it can be evaluated analytically in a non-perturbative manner by virtue of supersymmetry.   
Here the method used is called the supersymmetric localization \cite{Pestun:2016zxk}. 
Since it is quite difficult in general to calculate path-integrals or correlation functions analytically, 
one usually evaluate quantum fluctuations around a classical configuration where coupling constants are small enough to implement perturbative expansions. 
However, if a theory possesses supersymmetry, one can enjoy significant benefit from supersymmetry and the saddle-point method, 
which is the supersymmetric localization. 

In order to utilize the supersymmetric localization, a theory considered has to be invariant under supersymmetry. 
As such, there exist two approaches depending on how to realize supersymmetry on a curved compact space:  
One approach is to use supergravity theories \cite{Festuccia:2011ws, Dumitrescu:2012ha, Dumitrescu:2011iu, Komargodski:2010rb}, 
by which one can have a quite generic and systematic discussion, but there exist technical difficulties tied to the complexity of supergravity theories. 
The other is to use the iterative Weyl map \cite{Imamura:2011uw, Doroud:2012xw, Hama:2010av}. 
In this approach, one can construct a supersymmetric Lagrangian for a restricted class of manifolds that are conformally-flat,  
implementing supersymmetric transformations with transformation parameters satisfying the conformal Killing equation. 
In this article, we apply the latter approach.         
 
In the supersymmetric localization, configurations that contribute to a path-integral are localized only on configurations at the saddle-points, 
and fluctuation parts become one-loop exact. 
This is a powerfulness of the supersymmetric localization and the reason why it means a non-perturbative analysis; 
one can reduce infinite dimensional path-integrals to finite dimensional integrals or summations. 
Saddle-points in the supersymmetric localization are classified into two kinds, the Higgs branch and the Coulomb branch, whose partition functions are denoted by $Z_{\rm Higgs}$ and $Z_{\rm Coulomb}$, respectively.  
A vortex configuration is included in the Higgs branch since the defining equation of the saddle-point of the Higgs branch includes the vortex equation.
However, the partition function does not depend on the choice of the branch even though its appearance looks different, meaning that $Z_{\rm Higgs}=Z_{\rm Coulomb}$. 
In this article, we choose the Coulomb branch so as to avoid technical difficulties. 

The purpose of this article is to analyze quantum aspects of the Taub-NUT space which is the background of the KK$5$-branes from the string-theory point of view in a non-perturbative fashion. 
In the context of GLSM, the world-sheet instanton corrections to the single-centered Taub-NUT space was discussed in \cite{Harvey:2005ab}, 
and a physical interpretation of the world-sheet instantons was given in \cite{Okuyama:2005gx}.   
Namely, the article \cite{Okuyama:2005gx} discussed the generic multi-centered Taub-NUT space 
and interpreted the world-sheet instantons in the multi-centered case as the instantons originated with non-trivial $2$-cycles defined by pairs of the centers; 
through this interpretation, the world-sheet instantons in the single-centered case were identified with the disk instantons.           

The above discussion about the world-sheet instantons is performed restrictively in the $g\to 0$ limit  
where $g$ is the asymptotic radius of the $S^1$ fiber of the Taub-NUT space, 
although from the geometric point of view, the physical meaning of this limiting procedure would not be clear. 
In addition, for the single-centered case, since a map of the disk instanton introduced in \cite{Okuyama:2005gx} is well-defined for a finite value of $g$, 
it would be natural to consider that the disk instantons exist apart from $g=0$. 
Therefore, we wish to discuss the world-sheet instantons in a finite-$g$ region. 

For this purpose, the supersymmetric localization is quite useful since it allows us to evaluate the path-integral for arbitrary $g$. 
Therefore, in this article we analytically calculate the sphere partition function based on the supersymmetric localization from the standpoint of GLSM, 
from which we discuss the world-sheet instantons in a finite-$g$ region. 
As a result, we show that for the single-centered case, the world-sheet instantons exist only if $g = 0$\footnote{In the appendix \ref{sec:multi}, we calculate the sphere partition function for the multi-centered case in a finite-$g$ regime.}.

This article is organized as follows. 
In section \ref{sec:GLSM}, we give a brief introduction to GLSM for the KK$5$-branes on the flat space
in which we explain field contents, and then define GLSM for the KK$5$-branes on $S^2$ using the Weyl map from that on $\mbb{R}^2$.    
In section \ref{sec:Localization}, we overview ideas and a broad process of the supersymmetric localization, and perform the supersymmetric localization of GLSM for the KK$5$-branes on $S^2$ constructed in section \ref{sec:GLSM}.    
In section \ref{sec:metric}, we give an interpretation to the sphere partition function and discuss the world-sheet instantons appearing in the partition function.  
Section \ref{sec:metric} is devoted to summary and discussion.   
We also provide $6$ appendices to supplement the body.


\section{GLSM for KK5-branes}
\label{sec:GLSM}

In this section, we briefly review the gauged linear sigma model (GLSM) for the KK$5$-branes of which we wish to calculate the sphere partition function. 
This theory is constructed by a series of T-duality transformations of GLSMs \cite{Tong:2002rq, Harvey:2005ab, Kimura:2013fda, Okuyama:2005gx}.

\subsection{$U(1)^k$ GLSM for KK5-branes on $\mbb{R}^2$}
\label{sec:GLSM:2.1}

To begin with, we give the Lagrangian density of the $U(1)^k$ GLSM for multi-centered KK5-branes on the flat space \cite{Kimura:2013fda, Harvey:2005ab, Okuyama:2005gx}\footnote{Here although we obey a convention \cite{Kimura:2013fda}, in \cite{Harvey:2005ab, Okuyama:2005gx} gauge charges of chrged chiral multiplet $\wtilde{Q}, Q$ is fliped. } :
\bal
 \mc{L}_{\rm KK} 
 &=  \int d^4\theta \frac{1}{g^2} \ol{\Psi}\Psi 
 + \int d^4\theta \sum_{a=1}^{k}\left\{ \frac{1}{e_a^2}\bigl( -\ol{\Sigma}_a\Sigma_a + \ol{\Phi}_a\Phi_a \bigl) + \ol{Q}_a e^{-2V_a} Q_a + \ol{\wtilde{Q}}_a e^{+2V_a} \wtilde{Q}_a  \right\} \nn\\
 &\ \ \ + \int d^4\theta~  \frac{g^2}{2} \left( \sum_{a=1}^k 2V_a + \Gamma + \ol{\Gamma} \right)^2 
 + \sum_{a=1}^k \left\{ \sqrt{2}\int d^2\theta  \left( \wtilde{Q}_a\Phi_aQ_a + (s_a - \Psi)\Phi_a \right)+ (h.c.)\right\} \nn\\
 &\ \ \ + \sum_{a=1}^k \left\{ \sqrt{2}\int d^2\wtilde{\theta}\ t_a \Sigma_a + (h.c.) \right\}\ , 
 \label{eq:lkk}
\eal
where $a$ is a label for the $U(1)$ gauge group and $(h.c.)$ means Hermitian conjugate of the preceding terms. 
The Lagrangian density (\ref{eq:lkk}) consists of three parts: 
\bal 
 \mc{L}_{\rm KK} 
 =   \mc{L}_{\Sigma\Phi} 
 + \mc{L}_{Q \wtilde{Q}}
 + \mc{L}_{\Psi\Gamma}\ ,
 \label{eq:lkk2}
\eal 
where $\mc{L}_{\Sigma\Phi}$, $\mc{L}_{Q \wtilde{Q}}$ and $\mc{L}_{\Psi\Gamma}$ are Lagrangian densities 
for the $\mc{N}=(4,4)$ vector multiplet, charged hypermultiplet  and neutral hypermultiplet. 
In the following, we will explain each multiplet in detail.

The first one is the $\mc{N}=(4,4)$ vector multiplet that includes a vector field $A_m$, 
real and complex scalars $(\sigma, \eta, \phi)$ with $\sigma$ and $\eta$ being real scalars, 
and two Dirac fermions $(\lambda, \wtilde{\lambda})$.

In fact, the $\mc{N}=(4,4)$ vector multiplet can be decomposed into the $\mc{N}=(2,2)$ vector multiplet $(A_m, \lambda, \sigma, \eta, D) \in \Sigma$ 
and the $\mc{N}=(2,2)$ adjoint chiral multiplet $(\phi, \wtilde{\lambda}, D_\Phi) \in \Phi$ 
where $\Sigma$ is a twisted chiral representation of the vector multiplet; 
$D$ and $D_\Phi$ denote auxiliary fields in each multiplet.
These multiplets, i.e. $\Sigma$ and $\Phi$, have a vector-like $U(1)_R$ symmetry charge, $(q_\Sigma, q_\Phi) = (0,2)$. 
The $\mc{N}=(4,4)$ vector multiplet consists of the following Lagrangian densities: 
\bal
 \mc{L}_{\Sigma\Phi} ~: \ \ \ \int d^4\theta \frac{1}{e^2}\bigl( -\ol{\Sigma}\Sigma + \ol{\Phi}\Phi \bigl)\ , \ \ \  
 \int d^2\wtilde{\theta}~ t \Sigma\ , \ \ \  
 \int d^2\theta~ s \Phi\ , 
 \label{eq:lsigmaphi}
\eal
where $t := \frac{1}{\sqrt{2}}(t_1 + i t_2)$ and $s := \frac{1}{\sqrt{2}}(s_1 + is_2)$, 
and they are complex non-dynamical FI-parameters introduced as moduli in the Taub-NUT space \cite{Okuyama:2005gx}. 
These parameters specify the position of branes in the target space. 
The middle term in (\ref{eq:lsigmaphi}) is the twisted superpotential for the twisted chiral multiplet, 
and the last term in (\ref{eq:lsigmaphi}) the superpotential for the chiral multiplet. 
In this sense, non-dynamical FI-parameters, $t_1$ and $t_2$, are respectively the FI- and $\theta$-parameters, which form the complexified K\"ahler moduli. 
On the other hand, the parameters, $s_1$ and $s_2$, correspond to the other FI-parameters.

The $\mc{N}=(4,4)$ charged hypermultiplet is the second supermultiplet that involves two complex scalars $(q, \wtilde{q})$ and two Dirac fermions $(\psi, \wtilde{\psi})$, 
which can be decomposed into the $\mc{N}=(2,2)$ chiral multiplet $(q, \psi, F) \in Q$ with the $U(1)$ charge being $-1$ 
and the $\mc{N}=(2,2)$ chiral multiplet $(\wtilde{q}, \wtilde{\psi}, \wtilde{F}) \in \wtilde{Q}$ with the $U(1)$ charge being $+1$. 
Here $F$ and $\wtilde{F}$ are auxiliary fields in each multiplet. 
These multiplets, $Q$ and $\wtilde{Q}$, have the vector-like $U(1)_R$ charge, $(q_Q, q_{\wtilde{Q}}) = (0,0)$. 
The Lagrangian density of the $\mc{N}=(4,4)$ charged hypermultiplet consists of the two parts: 
\bal
  \mc{L}_{Q \wtilde{Q}} ~:~ \int d^4\theta \left\{ \ol{Q} e^{-2V} Q + \ol{\wtilde{Q}} e^{+2V} \wtilde{Q} \right\}\ ,  \ \ \  \int d^2\theta~ \wtilde{Q} \Phi Q\ . 
\label{eq:lqqtilde}  
\eal
This charged hypermultiplet has a $U(1)_f$ charge. 
Under the $U(1)_f$, the $\mc{N}=(2,2)$ multiplets, $\wtilde{Q} $ and $Q$, are respectively charged as $+1$ and $-1$.  
The last term in (\ref{eq:lqqtilde}) corresponds to a superpotential that makes the theory invariant under the R-symmetry
when extending the supersymmetry from $\mc{N}=(2,2)$ to $\mc{N}=(4,4)$.

The last supermultiplet is the $\mc{N}=(4,4)$ neutral hypermultiplet; 
it can be decomposed into the $\mc{N}=(2,2)$ chiral multiplet $(r^1, r^2, \chi, G)\in \Psi$ and the $\mc{N}=(2,2)$ St\"uckelberg-type chiral multiplet $(r^3, \gamma^4, \wtilde{\chi}, G_\Gamma)\in \Gamma$
\footnote{
This St\"uckelberg-type chiral multiplet $\Gamma$ shows up through the T-duality transformation from the GLSM for the H-monopole. 
This T-dulity is performed in the $S^1$ direction of the target space which is the imaginary part of the scalar component belonging to the twisted chiral $\Theta$. 
Therefore, besides the imaginary part of the scalar, components of $\Gamma$ are related to $\Theta$ by the duality relation: $ \Theta + \ol{\Theta} = - g^2( \Gamma + \ol{\Gamma} ) - 2g^2V$.}. 
Here $G$ and $G_\Gamma$ are auxiliary fields in each multiplet. 
These multiplets, $\Psi$ and $\Gamma$, have the vector like $U(1)_R$ charge $(q_\Psi, q_\Gamma) = (0,0)$. 
The $\mc{N}=(4,4)$ neutral hypermultiplet describes the target-space geometry, i.e. the Taub-NUT space, 
and its Lagrangian density consists of the following two parts: 
\bal
 \mc{L}_{\Psi\Gamma} ~:~ \int d^4\theta \left[ \frac{1}{g^2} \ol{\Psi}\Psi + \frac{g^2}{2} \biggl( 2V + \Gamma + \ol{\Gamma} \biggl)^2 \right], \hs{20pt} \int d^2\theta (- \Psi\Phi)\ ,
 \label{eq:lpsigamma}
\eal
where the coupling constant $g$ is interpreted as the asymptotic radius of the circle in the Taub-NUT space. 
The appearance of $g^2$ in front of the kinetic term of $\Gamma$ in (\ref{eq:lpsigamma}) implies that 
the model is obtained through the T-duality transformation of GLSM for the NS$5$-branes.

The GLSM we consider is the UV-theory of the NLSM with the KK$5$-brane background geometry.  
In fact, the low energy limit, $e \rightarrow \infty$, 
which makes massive fields decouple, yields the NLSM as an effective theory. 
We can read off the target-space metric $G_{mn}$ and the NS 2-form $B_{mn}$ from the world-sheet sigma model given by the following Lagrangian density:
\bal
 \mc{L}_{\rm NLSM} \ni - \frac12 G_{mn} \del_a X^m \del_b X^n h^{ab} + \frac12 B_{mn} \del_a X^m \del_b X^n \epsilon^{ab}\ ,
\eal 
where $h^{ab}$ and $\epsilon^{ab}$ are the metric and the epsilon tensor on the world-sheet.
GLSM for the KK5-brane has been obtained by the T-duality transformation of GLSM for the smeared NS5-brane in terms of not the component fields but the superfield description \cite{Tong:2002rq} (for the details, see \cite{Harvey:2005ab, Okuyama:2005gx, Kimura:2013fda}).

\subsection{GLSM for KK5-branes on $S^2$}
\label{sec:GLSM:2.2}

We wish to construct GLSM for the KK5-branes on $S^2$. In that regard, we first explain the supersymmetry we use.     
Concerning the supersymmetry transformation on $S^2$,
we have the conformal Killing spinor equations \cite{Benini:2012ui, Sugishita:2013jca, Closset:2014pda, Lu:1998nu}:
\bal
 \nabla_m\epsilon = \frac{i}{2r}\gamma_m\epsilon\ , \ \ \  
 \nabla_m\ol{\epsilon} = \frac{i}{2r}\gamma_m\ol{\epsilon}\ , 
 \label{eq:ckse}
\eal
where $\epsilon$ and $\ol{\epsilon}$ are parameters of the supersymmetry transformation.  
In the flat-space case, they are constant spinors, i.e. $\del_m\epsilon = \del_m\ol{\epsilon} = 0$. 
These equations constrain the parameters of the supersymmetry on $S^2$.

One can construct the supersymmetry transformation on a conformally flat manifold, 
deforming the transformation rule on $\mbb{R}^2$ in a Weyl covariant manner \cite{Imamura:2011uw, Doroud:2012xw, Benini:2015isa, Hama:2010av}.  
In this way, we obtain the supersymmetry transformation on $S^2$ \cite{Benini:2012ui}. 
For the $\mc{N}=(2,2)$ vector multiplet, we have 
\begin{align}
\begin{aligned}
 \delta \sigma_a &= \frac12 ( \ol{\epsilon}\lambda_a - \ol{\lambda}_a\epsilon )\ , \\
 \delta \eta_a &= - \frac{i}{2} ( \ol{\epsilon}\gamma^3\lambda_a - \ol{\lambda}_a\gamma^3\epsilon )\ , \\
 \delta A_{m,a} &= - \frac{i}{2} ( \ol{\epsilon}\gamma_m\lambda_a - \ol{\lambda}_a\gamma_m\epsilon )\ , \\
 \delta\lambda_a &= i \gamma_3\epsilon \Bigl( F_{\hat{1}\hat{2},a} - \frac{\eta_a}{r} \Bigl) - \epsilon \Bigl( D_a + \frac{\sigma_a}{r} \Bigl) + i \gamma^m \epsilon \nabla_m\sigma_a - \gamma_3\gamma^m\epsilon \nabla_m\eta_a\ , \\ 
 \delta \ol{\lambda}_a &= i \gamma_3\ol{\epsilon} \Bigl( F_{\hat{1}\hat{2},a} - \frac{\eta_a}{r} \Bigl) + \ol{\epsilon} \Bigl( D_a + \frac{\sigma_a}{r} \Bigl) - i \gamma^m \ol{\epsilon} \nabla_m\sigma_a - \gamma_3\gamma^m \ol{\epsilon} \nabla_m\eta_a\ , \\
 \delta D_a &= - \frac i2 ( \nabla_m \ol{\lambda}_a \gamma^m\epsilon + \ol{\epsilon}\gamma^m \nabla_m\lambda_a ) - \frac{1}{2r}( \ol{\epsilon} \lambda_a - \ol{\lambda}_a \epsilon)\ ,
\label{trans;vec}
\end{aligned}
\end{align}
where the suffix $a$ is a label for the $U(1)$ gauge group.
As well, concerning the $\mc{N}=(2,2)$ charged chiral multiplet with the $U(1)$ R-symmetry charge $q$, 
the supersymmetry transformation is given as follows: 
\begin{align}
\begin{aligned}
 \delta \phi_a &= \ol{\epsilon}\psi_a, \\
 \delta \ol{\phi}_a &= \ol{\psi}_a\epsilon, \\
 \delta \psi_a &= \Bigl( i\gamma^m D_m \phi_a + i \sigma\phi_a + \gamma_3\eta_a\phi_a - \frac{q}{2r}\phi_a \Bigl) \epsilon + \ol{\epsilon} F_a, \\
 \delta \ol{\psi}_a &= \Bigl( i\gamma^m D_m \ol{\phi}_a + i \ol{\phi}_a\sigma_a - \gamma_3\ol{\phi}_a\eta_a - \frac{q}{2r}\ol{\phi}_a \Bigl) \ol{\epsilon} + \ol{F}_a\epsilon, \\
 \delta F_a &= \epsilon \Bigl( i \gamma^m D_m\psi_a - i \sigma_a \psi_a + \gamma_3\eta_a \psi_a + \frac{q}{2r}\psi_a - i \lambda_a\phi_a \Bigl), \\
 \delta \ol{F}_a &= \ol{\epsilon} \Bigl( i \gamma^m D_m \ol{\psi}_a - i \ol{\psi}_a\sigma_a - \gamma_3 \ol{\psi}_a\eta_a + \frac{q}{2r}\ol{\psi}_a + i \ol{\phi}_a\ol{\lambda}_a \Bigl).
 \label{trans;chi}
\end{aligned}
\end{align}
where $D_m$ is the covariant derivative containing all connections. 
For instance, the scalar fields $(q, \wtilde{q}, \gamma)$ which belong to the chiral multiplets, $Q$, $\wtilde{Q}$ and $\Gamma$, are as follows:
\bal
 D_m q_a = \nabla_m q_a + i A_{m,a} q_a \ , \hs{10pt} D_m \wtilde{q}_a = \nabla_m q_a - i A_{m,a} q_a \ , \hs{10pt} D_m \gamma^4 = \nabla_m \gamma^4 + \sqrt{2} \sum_a A_{m,a} \ . 
\eal
The scalar $\gamma^4$ is an imaginary part of the lowest component $\frac{r^3 + i\gamma^4}{\sqrt{2}}$ in the St\"uckelberg-type chiral multiplet $\Gamma$.

Let us construct a supersymmetric Lagrangian on $S^2$, based on the approach \cite{Imamura:2011uw, Doroud:2012xw, Benini:2015isa, Hama:2010av}, 
in which one adds appropriate terms of $\mathcal{O}(1/r)$ and $\mathcal{O}(1/r^2)$ with $r$ being the radius of $S^2$ to the supersymmetric Lagrangian on $\mbb{R}^2$. 
As such, we start with the flat-space supersymmetric Lagrangian. 
The Lagrangian density on $\mbb{R}^2$ denoted by $\mc{L}_{\mbb{R}^2}$ is, by definition, invariant under the supersymmetry transformation on the flat space, $\delta_{\mbb{R}^2}$, i.e. 
\bal
 \mc{L}_{\mbb{R}^2}  \rightarrow  \mc{L}_{\mbb{R}^2} + \delta_{\mbb{R}^2}\mc{L}_{\mbb{R}^2}\ , \ \ \  \text{with}\ \ \ 
 \delta_{\mbb{R}^2}\mc{L}_{\mbb{R}^2} = 0\ .
\eal
Here we consider the supersymmetry transformation parameter satisfying the conformal Killing spinor equations.
Using this parameter we can obtain the supersymmetry transformation on $S^2$ (\ref{trans;vec}) and (\ref{trans;chi}).
Using the supersymmetry transformation on $S^2$ denoted by $\delta_{S^2}$, 
we formally transform the supersymmetric Lagrangian density on $\mbb{R}^2$: 
\bal
 \mc{L}_{\mbb{R}^2} \rightarrow \mc{L}_{\mbb{R}^2} + \delta_{S^2}\mc{L}_{\mbb{R}^2}\ .
\eal
Of course, the Lagrangian density on $\mbb{R}^2$ is not invariant under this supersymmetry transformation, 
but the point is that we can recast $\delta_{S^2}\mc{L}_{\mbb{R}^2}$ as  
\bal
\delta_{S^2}\mc{L}_{\mbb{R}^2} = \frac1r\delta_{S^2}(\cdots) + \frac1{r^2}\delta_{S^2}(\cdots)\ .
\eal
 We can define the Lagrangian density adding appropriate extra terms of $\mathcal{O}(1/r)$ and $\mathcal{O}(1/r^2)$ to $\mc{L}_{\mbb{R}^2}$ 
in such a way that the transformation $\delta_{S^2}$ of the extra terms exactly cancels $\delta_{S^2}\mc{L}_{\mbb{R}^2}$:  
\bal
 \mc{L}_{S^2} := \mc{L}_{\mbb{R}^2} - \frac1r(\cdots) - \frac1{r^2}(\cdots)\ . 
\eal
This procedure is well defined since $S^2$ is conformally flat. 
In this way, we can obtain the supersymmetric Lagrangian on $S^2$ that is, by definition, invariant under the supersymmetry transformation on $S^2$. 
Note that the method above using the conformal map may not be valid for a generic manifold and 
in that case, applying the supergravity approach, one can construct a Lagrangian of a supersymmetric gauge theory on a generic manifold \cite{Festuccia:2011ws, Dumitrescu:2012ha, Dumitrescu:2011iu, Komargodski:2010rb}.

Following the procedure above, we obtain the full Lagrangian density of GLSM for the KK5-branes on $S^2$ in the component-field formulation\footnote{We here apply the normalization of the Lagrangian in \cite{Benini:2012ui}.}:
\begin{align}
\begin{aligned}
\mc{L}_\Sigma &= \sum_{a=1}^k \left[ \frac12 \Bigl( F_{{\hat{1}\hat{2}}, a} - \frac{\eta_a}{r} \Bigl)^2 + \frac12\Bigl( D_a + \frac{\sigma_a}{r} \Bigl)^2 + \frac12(\nabla\sigma_a)^2 + \frac12(\nabla\eta_a)^2 + \frac i2 \ol{\lambda_a}\gamma^m\nabla_m\lambda_a \right]\ , \\
\mc{L}_\Phi &= \sum_{a=1}^k \left[ |\nabla\phi_a|^2 + |D_{\Phi,a}|^2 - i \ol{\wtilde{\lambda}}_a \gamma^m\nabla_m \wtilde{\lambda}_a - \frac{1}{r} \ol{\wtilde{\lambda}}_a\wtilde{\lambda}_a \right] \ , \\
\mc{L}_{\wtilde{Q}} &= \sum_{a=1}^k \Bigl[ |D_m\wtilde{q}_a|^2 + (\sigma^2_a + \eta^2_a )|\wtilde{q}_a|^2 + |\wtilde{F}_a|^2 + i D_a|\wtilde{q}_a|^2 \nn\\
			&\hs{50pt}+ i (\wtilde{q}_a\ol{\wtilde{\psi}}_a\lambda_a - \ol{\wtilde{q}}_a\ol{\lambda}_a\wtilde{\psi}_a) - i \ol{\wtilde{\psi}}_a\gamma^mD_m\wtilde{\psi}_a + i \ol{\wtilde{\psi}}_a( \sigma_a + i \gamma^3\eta_a )\wtilde{\psi}_a \Bigl] \ , \\
\mc{L}_{Q} &= \sum_{a=1}^k \Bigl[ |D_mq_a|^2 + (\sigma^2_a + \eta^2_a )|q_a|^2 + |F_a|^2 - i D_a|q_a|^2 \nn\\
			&\hs{50pt}- i (q_a\ol{\psi}_a\lambda_a - \ol{q}_a\ol{\lambda}_a\psi_a) - i \ol{\psi}_a\gamma^mD_m\psi_a - i \ol{\psi}_a( \sigma_a + i \gamma^3\eta_a )\psi_a \Bigl] \ , \\
\mc{L}_\Gamma &= \frac1{g^2} \Bigl[ \frac12 (\nabla_m r^3)^2 - i \ol{\wtilde{\chi}} \gamma^m \nabla_m\wtilde{\chi} \Bigl] + g^2 \Bigl[ \frac12(D_m\gamma^4)^2 + |G_\Gamma|^2 \Bigl] \\
 					&\hs{50pt}+ \frac{g^2}2 \sum_{a,b=1}^k ( \sigma_a\sigma_b + \eta_a\eta_b ) + i\sqrt{2}r^3 \sum_{a=1}^k D_a + i \sum_{a=1}^k \Bigl[ \ol{\wtilde{\chi}}\gamma_3\lambda_a + \ol{\lambda}_a\gamma_3\wtilde{\chi} \Bigl]\ , \\
\mc{L}_\Psi &= \frac12(\nabla_mr^1)^2 + \frac12(\nabla_mr^2)^2 + |G|^2 - i \ol{\chi}\gamma^m\nabla_m \chi\ ,
\end{aligned}
\end{align}

and 
\begin{align}
\begin{aligned}
\mc{L}_{\wtilde{Q}\Phi Q} &= i \sqrt{2} \sum_{a=1}^k \Bigl[ \phi_a ( q_a \wtilde{F}_a + \wtilde{q}_aF_a ) + \ol{\phi}_a ( \ol{q}_a \ol{\wtilde{F}_a } + \ol{\wtilde{q}}_a \ol{F}_a ) + q_a\wtilde{q}_aD_{\Phi,a} + \ol{q}_a\ol{\wtilde{q}}_a\ol{D}_{\Phi,a} \Bigl] \\
 				&\hs{40pt}- i \sqrt{2} \sum_{a=1}^k \Bigl[ (\phi_a \wtilde{\psi}_a \psi_a + \ol{\phi}_a \ol{\wtilde{\psi}}_a \ol{\psi}_a ) + ( q_a \wtilde{\lambda}_a\wtilde{\psi}_a + \ol{q}_a\ol{\wtilde{\lambda}}_a\ol{\wtilde{\psi}}_a ) + (\wtilde{q}_a\wtilde{\lambda}_a\psi_a + \ol{\wtilde{q}}_a\ol{\wtilde{\lambda}}_a\ol{\psi}_a ) \Bigl]\ , \\[10pt]
 \mc{L}_{\Psi\Phi} &= i \sum_{a=1}^k \Bigl[ \sqrt{2}\phi_a G + \sqrt{2}\ol{\phi}_a \ol{G} + (r^1 + i r^2)D_{\Phi,a} + (r^1 - i r^2)\ol{D}_{\Phi,a} - \sqrt{2} \chi\wtilde{\lambda}_a - \sqrt{2} \ol{\chi}\ol{\wtilde{\lambda}}_a \Bigl]\ , \\[10pt]
 \mc{L}_{s\Phi} &= - i \sqrt{2} \sum_{a=1}^k \Bigl[ s_a D_{\Phi,a} + \ol{s}_a \ol{D}_{\Phi,a} \Bigl]\ , \hs{30pt}
\mc{L}_{t\Sigma} = - i \sqrt{2} \sum_{a=1}^k \Bigl[ t_{1,a}D_a - t_{2,a} F_{ {\hat{1}\hat{2}},a} \Bigl]\ . 
\label{eq:lagrangian}
\end{aligned}
\end{align}
where $(a,b)$ are labels for the $U(1)$ gauge group, $r$ is the radius of $S^2$, 
and the kinetic term of the St\"uckelberg-type chiral multiplet $\Gamma$ is defined by $D_m\gamma^4 = \nabla_m \gamma^4 + \sqrt{2}\sum_a A_{m,a}$.
The Lagrangian above respects the $\mc{N}=(2,2)$ supersymmetry, while the $\mc{N}=(4,4)$ supersymmetry is not manifest 
because the $SU(2)_R$ symmetry is not manifest in the Lagrangian. 
One can check that the theory can restore the full supersymmetry in the flat space limit, $r \rightarrow \infty$. 
Since the supersymmetry is conserved, we can perform the supersymmetric localization.

\section{Localization}
\label{sec:Localization}

\subsection{Procedure}
\label{sec:Procedure}

We overview basic concepts of the supersymmetric localization \cite{Pestun:2007rz}. 
Namely, we provide no details about calculations to place a priority on sketching the whole story of supersymmetric localization (for a comprehensive review, see e.g.  \cite{Pestun:2016zxk}).

The supersymmetric localization is a powerful technique that allows us to implement the path-integral of supersymmetric field theory exactly: 
If supersymmetry $\mc{Q}$ exists as a symmetry of the theory we consider and if it is conserved even at the quantum level, 
one can deform the original action $S$ by a $\mc{Q}$-exact term $\mc{Q}V$ as $S \rightarrow S + t \mc{Q}V$. 
With tuning $t \to \infty$, the path-integral can be localized at the field configurations satisfying $\mc{Q}V=0$.

In the following, we clarify the statement above. Let us consider a supersymmetric theory with a supersymmetry $\mc{Q}$ that is preserved quantum-mechanically, 
whose partition function is given as 
\bal
 Z_0 = \int \mc{D}\Phi \ e^{- S[\Phi]} \ ,
 \label{eq:z0}
\eal
where $\Phi$ correctively denotes all fields in the theory.

We wish to deform the partition function (\ref{eq:z0}) by a $\mc{Q}$-exact term. Since by definition, $\mc{Q}$ is a symmetry of the theory, we have $\mc{Q} S = 0$. 
We can find $\mc{Q}V$ such that $\mc{Q}^2V=0$ and the bosonic part of $\mc{Q}V$ is semi-positive definite, 
through which we can deform the partition function (\ref{eq:z0}) by adding $\mc{Q}V$ to the action:       
\bal
 Z[t] = \int \mc{D}\Phi \ e^{- S[\Phi] - t \mc{Q}V[\Phi]} \ , 
 \label{eq:zt}
\eal
where $t$ is an arbitrary parameter. 
The important point here is that $Z[t]$ is, in fact, independent of $t$. 
This can be checked differentiating the deformed partition function (\ref{eq:zt}) w.r.t. $t$, which yields    
\bal
 \frac{d}{dt} Z[t] = \int \mc{D}\Phi \ (\mc{Q}V) \ e^{- S - t \mc{Q}V} = \int \mc{D}\Phi \ \mc{Q} \left( V \ e^{- S - t \mc{Q}V} \right) = 0 \ ,
\eal
where we have used the anomaly-free condition, $\mc{Q}(\mc{D}\Phi)=0$. 
Therefore, the following relation holds:   
\bal
 Z_0 
 = \lim_{t\to \infty} Z[t]
 =: Z_{\infty}
 \ , 
\eal
which means that the original theory $Z_0$ can be precisely evaluated in terms of $Z_{\infty}$. 
Although it is often difficult to perform the original path-integral $Z_0$, if the supersymmetry is quantum-mechanically conserved, 
one can evaluate $Z_{\infty}$ instead, which is much easier. 
From (\ref{eq:zt}), the path-integral $Z_{\infty}$ is indeed localized at the saddle-point defined by $\mc{Q}V=0$.

Next, we discuss the saddle-point method for the deformed path-integral (\ref{eq:zt}) at large $t$. 
As mentioned, the saddle-point is defined by    
\bal
 \mc{Q}V[\Phi] = 0\ .
 \label{eq:qvis0}
\eal
If denoting by $\Phi_0$ the field configuration satisfying (\ref{eq:qvis0}), we expand all the fields in the theory around $\Phi_0$: 
\bal
 \Phi = \Phi_0 + \frac{1}{\sqrt{t}} \what{\Phi}\ , 
 \label{eq:expandphi}
\eal
where $\what{\Phi}$ is a fluctuation around the saddle-point configuration. 
Inserting (\ref{eq:expandphi}) into (\ref{eq:zt}), one obtains 
\bal
 Z[t] = \int \mc{D}\Phi \ e^{- S[\Phi] - t \mc{Q}V[\Phi]} 
 \overset{t\rightarrow\infty}{=} 
 \int \mc{D}\Phi_0 \ e^{- S[\Phi_0]} \ Z_{\text{$1$-loop}}[\Phi_0]\ . 
 \label{eq:zt2}
\eal
Here $Z_{\text{$1$-loop}}$ means the $1$-loop contributions that appear due to Gaussian integrals associated with quadratic terms in $\mc{Q}V[\Phi]$. 
Concerning the contributions from higher loops, they vanish at large $t$, i.e. $1$-loop exact. Thus the localization can treat non-perturbative effects.  
This is a powerfulness of the localization method. Based on the localization, one can reduce path-integrals that generally include infinite-dimensional integrals to finite-dimensional integrals or summations. 

We have briefly sketched some basic properties of the localization method, although it is rather formal. 
Once we can find a suitable $\mc{Q}V$ term, what we need to do is to evaluate the $1$-loop determinants at large $t$.

We will conclude this subsection by commenting on the $1$-loop determinants $Z_{\text{$1$-loop}}$ that appear in (\ref{eq:zt2}). 
We need to implement Gaussian integrals such as 
\bal
 \int \mc{D}\phi\mc{D}\ol{\phi} \ e^{ - \int d^nx \sqrt{g} |\nabla_m\phi|^2} &\simeq \frac{1}{\det (-\nabla^2)}, \\
 \int \mc{D}\psi\mc{D}\ol{\psi} \ e^{ - \int d^nx \sqrt{g} i\ol{\psi}\gamma^m\nabla_m\psi} &\simeq \det (-i\gamma^m\nabla_m)\ , 
\eal
where $\phi$ and $\psi$ are bosonic and fermionic fields, respectively. 

Eigenvalues of each operator are necessary to evaluate the determinants.  
Here we consider the quantity $|\nabla_m\phi|^2$ defined on $S^2$, as an example. 
Since fields on $S^2$ can be expanded in terms of the spherical harmonics $Y_{jm}(\theta, \phi)$ (see appendix \ref{sec:harmonics}), 
we can find the eigenvalues of the Klein-Gordon operator:  
\bal
 - \nabla^2 Y_{jm}(\theta, \phi) = \frac{ j(j+1) }{r^2} Y_{jm}(\theta, \phi)\ . 
\eal
Here we have considered the gauge-neutral Lorentz scalar part as the simplest example.  
Concerning other fields, one can basically deal with their $1$-loop determinants in the similar manner, 
with special care for gauge charges and behaviors under the Lorentz transformation, if necessary.

\subsection{Deformation terms}
\label{sec:Deformation}

We employ the supersymmetric localization with the supercharge \cite{Benini:2012ui}: 
\bal
 \mc{Q} := Q + Q^\dagger, \hs{20pt} Q := \epsilon^\alpha Q_\alpha, \hs{20pt} Q^\dagger := {\epsilon^c}^\alpha Q^\dagger_\alpha \ . 
\eal
The operator $\mc{Q}$ is Grassmann odd; 
$\epsilon^c := C \epsilon^*$ where $\epsilon$'s are solutions of the conformal Killing spinor equations;
barred spinors defined as $\ol{\psi} = C(\psi^\dag)^T$; 
$\ol{\phi}= \phi^\dag$ (for details, see appendix \ref{sec:Supercharges}). 
Also we set $\epsilon^\dag \epsilon = 1$ and $\psi^\dag$ is contracted with $\psi$ such that $\psi^\dag \psi = \psi^\dag C \psi$ and $\psi^\dag \gamma^m \psi = \psi^\dag C \gamma^m\psi$.

Let us introduce the $\mc{Q}$-exact deformation terms, $\mc{Q}V$. 
Concerning the vector multiplet and the general chiral multiplet, 
the deformation terms discussed in \cite{Sugishita:2013jca} are 
\bal
 \mc{Q}V_{\rm vector} &= \frac14\mc{Q} \Bigl[ (\ol{\mc{Q}\lambda})\lambda + \lambda^\dagger(\ol{\mc{Q}\lambda^\dagger}) \Bigl]\ , \nn\\
 \mc{Q}V_{\rm chiral} &= \frac12\mc{Q} \Bigl[ (\ol{\mc{Q}\psi})\psi + \psi^\dagger(\ol{\mc{Q}\psi^\dagger}) \Bigl] + \frac{q-1}{2r} \mc{Q} \left[ \phi^\dag(\ol{\mc{Q}\phi^\dag}) - (\ol{\mc{Q}\phi})\phi \right]  \ .
\eal
Here the bar implies the hermitian conjugate for a c-number and transforms a field as $\Phi\rightarrow\Phi^{\dagger}$ such that $\ol{ (\Phi^\dag) } = \Phi$.

In the following, we enumerate the details: 
For the vector multiplet $(\sigma, \eta, \lambda, A, D)$, we have  
\bal
 \mc{Q}V^b_{\rm vector} &= \frac12(\epsilon^\dag\epsilon) \Bigl( F_{\hat{1}\hat{2}} - \frac{\eta}{r} \Bigl)^2 + \frac12(\epsilon^\dag\epsilon) \Bigl( D + \frac{\sigma}{r} \Bigl)^2 + \frac12(\epsilon^\dag\epsilon) (\nabla_m\sigma)^2 + \frac12(\epsilon^\dag\epsilon) (\nabla_m\eta)^2\ , \label{eq:QVvectorboson} \\[10pt]
 \mc{Q}V^f_{\rm vector} &= \frac i4 (\nabla_m\lambda^\dag\gamma^m\lambda)(\epsilon^\dag\epsilon) -  \frac i4(\lambda^\dag\gamma^m\nabla_m\lambda )(\epsilon^\dag\epsilon)\ ,
 \label{eq:QVvectorfermion}
\eal
where the suffixes, $b$ and $f$, specify the bosonic and fermionic parts, respectively, and hereafter we will use the suffixes in this sense. 
For the charged chiral multiplet $(\phi, \psi, F)$  with a generic R-charge $q$, we have 
\bal 
 \mc{Q}V_\psi := \frac12\mc{Q} \Bigl[ (\ol{\mc{Q}\psi})\psi + \psi^\dagger(\ol{\mc{Q}\psi^\dagger}) \Bigl]\ , \hs{20pt}
 \mc{Q}V_\phi := \mc{Q} \left[ \phi^\dag(\ol{\mc{Q}\phi^\dag}) - (\ol{\mc{Q}\phi})\phi \right]\ ,
\eal
where 
\bal
 \mc{Q}V^{b}_\psi &= (\epsilon^\dag\epsilon) \Bigl[ |D_m\phi|^2 + \phi^\dag\sigma^2\phi + \phi^\dag\eta^2\phi + \frac{q^2}{4r^2}\phi^\dag\phi + |F|^2 \Bigl] \nn\\
 							&\hs{20pt} + i\frac{1-q}{r} (\epsilon^\dag\gamma^m\epsilon) \phi^\dag D_m \phi + (\epsilon^\dag\gamma^3\epsilon) \Bigl[ \frac{2-q}{r} \phi^\dag\eta\phi - \phi^\dag F_{12}\phi \Bigl] - i(\epsilon^\dag\gamma^3\gamma^m\epsilon)\phi^\dag D_m\eta \phi \ ;
\eal
\bal
 \mc{Q}V^f_\psi &= (\epsilon^\dag\epsilon) \Bigl[ \frac{i}{2}D_m \psi^\dag \gamma^m\psi - \frac{i}{2}\psi^\dag\gamma^mD_m\psi + i \psi^\dag\sigma\psi - \psi^\dag\eta\gamma^3\psi - \frac{3i}{4}\phi^\dag \lambda^\dag \psi + \frac{3i}{4}\psi^\dag\lambda\phi - \frac{1}{2r}\psi^\dag\psi \Bigl] \nn\\
 					&\hs{20pt} + i (\epsilon^\dag\gamma^m\epsilon) \Bigl[ \frac{i}{2}\varepsilon_{mn} ( D^n\psi^\dag\gamma^3\psi + \psi^\dag\gamma^3D^n\psi) + \frac14\phi^\dag\lambda^\dag\gamma_m\psi - \frac14\psi^\dag \gamma_m\lambda\phi - \frac{iq}{2r}\psi^\dag\gamma_m\psi \Bigl] \nn\\
					&\hs{20pt} + i (\epsilon^\dag\gamma^3\epsilon) \Bigl[ \frac{1}{2}D_m\psi^\dag\gamma^3\gamma^m\psi + \frac{1}{2}\psi^\dag\gamma^3\gamma^mD_m\psi + \frac14\phi^\dag\lambda^\dag\gamma_3\psi - \frac14\psi^\dag \gamma_3\lambda\phi - \frac{i(q+1)}{2r}\psi^\dag\gamma_3\psi \Bigl] \nn\\
					&\hs{20pt} - \frac{i}{4} (\epsilon\gamma^m\epsilon) \psi^\dag\gamma_m\lambda^\dag\phi + \frac{i}{4}(\epsilon^\dag\gamma^m\epsilon^c)\phi^\dag\lambda\gamma_m\psi - \frac{i}{4} (\epsilon\gamma^3\epsilon) \psi^\dag\gamma_3\lambda^\dag\phi + \frac{i}{4}(\epsilon^\dag\gamma^3\epsilon^c)\phi^\dag\lambda\gamma_3\psi\ ; 
\label{check}
\eal
\bal
 \mc{Q}V^b_\phi 
 													&= 2(\epsilon^\dag\epsilon) \Bigl( i \phi^\dag \sigma\phi - \frac{q}{2r} \phi^\dag\phi \Bigl) + 2i ( \epsilon^\dag \gamma^m\epsilon) \phi^\dag D_m\phi + 2 ( \epsilon^\dag\gamma^3\epsilon) \phi^\dag\eta\phi\ ; \\
 \mc{Q}V^f_\phi  
 						&= - ( \epsilon^\dag \epsilon) ( \psi^\dag \psi) + i( \epsilon^\dag\gamma_3\epsilon) i( \psi^\dag\gamma^3\psi) + i( \epsilon^\dag\gamma_m\epsilon) i( \psi^\dag\gamma^m\psi)\ . 
\eal

The $\mc{Q}V$ term for the chiral multiplet becomes  
\bal
 \mc{Q}V^b_{\rm chiral}
							&= (\epsilon^\dag\epsilon) \Bigl[ |D_m\phi|^2 + \phi^\dag\sigma^2\phi + \phi^\dag\eta^2\phi + i\frac{q-1}{r}\phi^\dag\sigma\phi - \frac{q(q-2)}{4r^2}\phi^\dag\phi + |F|^2 \Bigl] \nn\\
							&\hs{20pt} + (\epsilon^\dag\gamma^3\epsilon) \phi^\dag\Bigl( \frac{\eta}{r} - F_{\hat{1}\hat{2}} \Bigl)\phi - i(\epsilon^\dag\gamma^3\gamma^m\epsilon)\phi^\dag (D_m\eta) \phi\ ;  
\label{eq:QVchiralboson}\\ 
  \mc{Q}V^f_{\rm chiral}
								&= ( \epsilon^\dag\epsilon) \Bigl[ \frac{i}{2}D_m\psi^\dag \gamma^m\psi - \frac{i}{2}\psi^\dag\gamma^mD_m\psi + i \psi^\dag\sigma\psi - \psi^\dag\eta\gamma^3\psi - \frac{3i}{4}\phi^\dag \lambda^\dag \psi + \frac{3i}{4} \psi^\dag\lambda\phi - \frac{q}{2r}\psi^\dag\psi \Bigl] \nn\\
 					&\hs{20pt} + i ( \epsilon^\dag \gamma^m\epsilon ) \Bigl[ \frac14 \phi^\dag \lambda^\dag \gamma_m\psi - \frac14 \psi^\dag \gamma_m\lambda\phi \Bigl] + i ( \epsilon^\dag\gamma^3\epsilon ) \Bigl[ \frac14 \phi^\dag \lambda^\dag \gamma_3\psi - \frac14 \psi^\dag \gamma_3\lambda\phi \Bigl] \nn\\
					&\hs{20pt} - \frac{i}{4} (\epsilon\gamma^m\epsilon) ( \psi^\dag\gamma_m\lambda^\dag)\phi + \frac{i}{4}( \epsilon^\dag \gamma^m \epsilon^c ) \phi^\dag (\lambda\gamma_m\psi) - \frac{i}{4} (\epsilon\gamma^3\epsilon) ( \psi^\dag\gamma_3\lambda^\dag)\phi + \frac{i}{4}( \epsilon^\dag \gamma^3 \epsilon^c ) \phi^\dag (\lambda\gamma_3\psi)\ . 
\label{eq:QVchiralfermion}
\eal

Using the $\mc{Q}V$-terms constructed above, 
we can evaluate the saddle-point of the supersymmetric localization. 
In particular, we choose the Coulomb branch configuration satisfying 
\bal
 F_{\hat{1}\hat{2}} = \frac{\eta}{r}\ , \hs{20pt} D = - \frac{\sigma}{r}\ , \hs{20pt} \sigma = \eta = {\rm constant}\ , \hs{20pt} \lambda = 0  \ ; 
 \label{eq:saddle1}
\eal
\bal
 \phi = \psi = F = 0 \ .
 \label{eq:saddle2}
\eal

\subsection{One-loop determinants}
\label{sec:determinant}

We calculate the one-loop determinants for all the multiplets in GLSM for the KK$5$-branes. 
In that regard, we follow the procedure discussed in \cite{Benini:2012ui}.\\

\noindent
\ul{\large Neutral chiral multiplet $\Psi$}\\

The relevant $\mc{Q}V$-term is  
\bal
 \mc{Q}V_\Psi = \frac12(\nabla_mr^1)^2 + \frac12(\nabla_mr^2)^2 - i \chi^\dag\gamma^m\nabla_m \chi + |G|^2 \ .
 \label{eq:qvpsi}
\eal
Using (\ref{eq:qvpsi}) one finds that the corresponding saddle-point values are constants that are set to be zero.     
The expansions around the saddle-point values are given as follows:  
\bal
 r = \frac{1}{\sqrt{t}}\what{r}\ , \hs{20pt} \chi = \frac{1}{\sqrt{t}}\what{\chi} \ ,
\eal
where $r := \frac{1}{\sqrt{2}}(r^1+ir^2)$.  
We can carry out the corresponding Gaussian integrals in terms of the spherical harmonics, 
which gives a constant. 
The $1$-loop determinant for the neutral chiral multiplet is trivial:
\bal
 Z_{\Psi} &= {\rm constant} \ .
\eal

Here $r$ and $\ol{r}$ include zero modes, $r_0$ and $\ol{r}_0$, 
and the zero-mode integrals are performed separately in the presence of the superpotential term:
\bal
 \int dD_{\Phi, 0}\  d\ol{D}_{\Phi, 0}\ dr_0\ d\ol{r}_0 \  e^{i \sqrt{2} (r_0 - s) D_{\Phi, 0} + i \sqrt{2} (\ol{r}_0 - \ol{s}) \ol{D}_{\Phi,0} - |D_{\Phi,0}|^2} = 1\ , 
\eal
where irrelevant numerical factors are absorbed into the definition of integral measures, 
and we respectively denote by $D_{\Phi, 0}$ and $\ol{D}_{\Phi, 0}$ the constant modes of the auxiliary field in the adjoint chiral multiplet.

The $|G|^2$-term in (\ref{eq:qvpsi}) will be treated in due course.  
\\ 

\noindent
\ul{\large Adjoint chiral multiplet $\Phi$}\\

The $\mc{Q}V$-term of the adjoint chiral multiplet includes 
\bal
\mc{Q}V_\Phi = |\nabla\phi|^2 - i \wtilde{\lambda}^\dag \gamma^m\nabla_m \wtilde{\lambda} - \frac{1}{r} \wtilde{\lambda}^\dag\wtilde{\lambda} + |D_{\Phi}|^2 \ .
 \label{eq:qvphi}
\eal
From (\ref{eq:qvphi}), the saddle-point values are determined to be constants that are taken to be zero. 
The expansions of the fields around the saddle-point values are given as 
\bal
 \phi = \frac{1}{\sqrt{t}}\what{\phi}, \hs{20pt} \wtilde{\lambda} = \frac{1}{\sqrt{t}}\what{\wtilde{\lambda}}\ , \hs{20pt} D_\Phi = \frac{1}{\sqrt{t}}\what{D}_\Phi \ .
\eal
Since the scalar is gauge neutral, the corresponding operator is a usual Laplacian, 
and we can immediately get the determinant.

The scalar has zero modes, $\phi_0$ and $\ol{\phi}_0$, that correspond to $j=0$, 
and the associated integral is performed with the zero modes of the auxiliary field in the neutral chiral multiplet, $G_0$ and $\ol{G}_0$:
\bal
 e^{ i\sqrt{2}\phi_0 G_0 + i \sqrt{2}\ \ol{\phi}_0 \ol{G}_0 - |G_0|^2 }
 \ ,
\eal
which gives a constant. 

The operator can be expressed as a matrix form\footnote{We have chosen the standard position of the spinor indices in the definition of the matrix \cite{Benini:2012ui}.} :
\bal
 \mc{Q}V_\Phi \ni - i \wtilde{\lambda}^\dag \gamma^m\nabla_m \wtilde{\lambda} - \frac{1}{r} \wtilde{\lambda}^\dag\wtilde{\lambda} =:
 \wtilde{\lambda}^\dag \mc{O}_{\wtilde{\lambda}} \wtilde{\lambda}\ , 
\label{eq:qvphi} 
\eal
where 
\begin{equation}
\mc{O}_{\wtilde{\lambda}} = 
\begin{pmatrix}
-\frac{1}{r} & -2i \nabla_+ \\
2i \nabla_- & + \frac{1}{r}
\end{pmatrix}\ .
\label{eq:olambda}
\end{equation}
The operator (\ref{eq:olambda}) acts on the spin spherical harmonics, $Y^{\frac12}_{j,m}$ and $Y^{-\frac12}_{j,m}$.  
Since the adjoint chiral multiplet is gauge singlet under the $U(1)$-gauge symmetry, 
the $1$-loop determinant becomes, for $\frac12 \le j$,  
\bal
 {\rm Det}' \ \mc{O}_{\wtilde{\lambda}} 
 							= \prod_{j=\frac32} \left( j + \frac32 \right)^{2j+1} \left( j - \frac12 \right)^{2j+1}\ ,
\eal
where the prime denotes to remove the zero-mode. 

The result for the adjoint chiral multiplet becomes
\bal
 Z_{\Phi} &= {\rm constant} \ .
 \label{eq:phi}
\eal
Note that there exist the fermionic zero-modes corresponding to $j=\frac12$, and therefore,  
we need to define the path integral measures of $\wtilde{\lambda}_0$ and $\wtilde{\lambda}^\dag_0$ properly in such a way that (\ref{eq:phi}) does not vanish.\\

\noindent
\ul{\large Charged chiral multiplets $\wtilde{Q}$, $Q$}\\

We will focus only on the charged chiral multiplet $\wtilde{Q}$, 
since the calculation associated with $Q$ makes no difference except for the sign-difference in the gauge charge. 
The $\mc{Q}V$-term for the charged chiral multiplet $\wtilde{Q}$ at the saddle-point is 
\bal
 \mc{Q}V_{\wtilde{Q}} &= | D_m\wtilde{q} |^2 + (\sigma^2_0 + \eta^2_0 )| \wtilde{q}|^2 - \frac{i\sigma_0}{r} | \wtilde{q}|^2 + |\wtilde{F}|^2 - i \wtilde{\psi}^\dag \gamma^mD_m\wtilde{\psi} + i \wtilde{\psi}^\dag( \sigma_0 + i \gamma^3\eta_0 )\wtilde{\psi}\ .
\eal
The expansions of fields in the multiplet are 
\bal
\wtilde{q}  = \frac{1}{\sqrt{t}}\what{\wtilde{q}}, \hs{20pt} \wtilde{\psi} = \frac{1}{\sqrt{t}}\what{\wtilde{\psi}}, \hs{20pt} \wtilde{F} = \frac{1}{\sqrt{t}}\what{\wtilde{F}} \ .
\eal
Hereafter we will omit the hat symbol for a notational simplicity. 
The integral in which $\wtilde{F}$ is involved is a Gaussian integral and it can be performed easily. 
Let us extract the operator associated with $\wtilde{q}$ and denote it by $\wtilde{q}^\dag\mc{O}_{\wtilde{q}}\wtilde{q}$:
\bal
 \mc{O}_{\wtilde{q}} = - D_m D^m - \frac{i\sigma_0}{r} + \sigma^2_0 + \eta^2_0 \ ,  
\eal
where we have used the fact that this multiplet has the zero R-charge, $q=0$. 
Since this operator acts on the scalar, the effective spin is given by $s_{\rm eff} = 0 - \frac{m}{2}$ (see appendix \ref{sec:harmonics} for the detail). 
From the saddle-point for the vector multiplet, we have $F_{\hat{1}\hat{2}} = \frac{\eta_0}{r} = \frac{m}{2r^2}$ and $m \in \mbb{Z}$.
Taking a multiplicity into account, the $1$-loop determinant becomes 
\bal
 \mbox{\rm Det} \ \mc{O}_{\wtilde{q}} &= \prod^{\infty}_{j=\frac{|m|}{2}} \Bigl( j - i\sigma_0 \Bigl)^{2j+1} \Bigl( j+1 + i\sigma_0 \Bigl)^{2j+1}\ . 
\eal

In the same way, we can extract the operator associated with $\wtilde{\psi}$ and denote it by $\mc{O}_{\wtilde{\psi}}$, 
which can be recast into a matrix form: 
\begin{align}
 \mc{O}_{\wtilde{\psi}} = \frac1r \left( \begin{array}{cc} i\sigma_0 - \frac{m}{2} &  - 2irD_+  \\  2irD_-  & - i\sigma_0 - \frac{m}{2} \end{array} \right)\ ,
\end{align}
where $D_\pm:=(D_1 \mp iD_2)/2$ \cite{Benini:2012ui}. 
Depending on the value of $j$, let us do the case study:
\begin{enumerate}

\item $\frac12 + \frac{|m|}{2} \le j$

In this case, for a given $(j,m)$ we have 
\begin{align}
 \mc{O}_{\wtilde{\psi}} = \frac1r \left( \begin{array}{cc} i\sigma_0 - \frac{m}{2}  &  - 2irD_+  \\  2irD_-  &  -i\sigma_0 - \frac{m}{2}  \end{array} \right) \ . 
\end{align}
where the fermion field is expanded by the spinor spherical harmonics.
The spin spherical harmonics which act the operator $\mc{O}_{\wtilde{\psi}}$ are $Y^{\frac12 - \frac{m}{2}}_{j,m}, Y^{-\frac12 - \frac{m}{2}}_{j,m}$.
The determinant of the operator $\mc{O}_{\wtilde{\psi}}$ becomes 
\begin{align}
 {\rm Det} ~ \mc{O}_{\wtilde{\psi}}  &= \prod_{j=\frac{|m|}{2}+\frac12}^\infty\Bigl( j + \frac{1}2 - i\sigma_0 \Bigl)^{2j+1} \Bigl( j + \frac{1}2 + i\sigma_0 \Bigl)^{2j+1}. 
\end{align}

\item $j=\frac{|m|}{2} - \frac12$

In this case, the operator have a rank $1$. 
Taking a multiplicity into account, the $1$-loop determinant becomes 
\bal
 {\rm Det} ~\mc{O}_{\wtilde{\psi}}  &= (-1)^{ \lfloor m \rfloor } \biggl( \frac{|m|}{2} -i\sigma_0 \biggl)^{|m|} .
\eal
where $\lfloor m \rfloor$ is the Gauss function. 
\end{enumerate}

From the case study above, 
the total $1$-loop determinant of the fermion becomes 
\bal
 {\rm Det} ~\mc{O}_{\wtilde{\psi}} &= \prod_{k=\frac{|m|}{2}}^\infty \Bigl( k - i\sigma_0 \Bigl)^{2k} \Bigl( ( k + 1 ) + i\sigma_0 \Bigl)^{2k+2}. 
\eal

\noindent
Wrapping up the discussion above, the $1$-loop determinant for the the charged chiral multiplets, $\wtilde{Q}$ ($Q=+1, q=0$) and $Q$ ($Q=-1, q=0$), become 
\bal
 Z_{\wtilde{Q}} = \frac{ \Gamma\Bigl( - \frac{m + n_f}{2} - i(\sigma_0 + \sigma_f ) \Bigl) }{ \Gamma\Bigl(1 - \frac{m + n_f}{2} + i(\sigma_0 + \sigma_f ) \Bigl)}, \hs{30pt}
 Z_{Q} = \frac{ \Gamma\Bigl( \frac{m + n_f}{2} + i(\sigma_0 + \sigma_f ) \Bigl) }{ \Gamma\Bigl(1 + \frac{m + n_f}{2} - i(\sigma_0 + \sigma_f ) \Bigl)}\ , 
\eal
where $n_f$ and $\sigma_f$ are twisted masses for the flavor symmetry, and they are introduced to regularize the $1$-loop determinant \cite{Benini:2012ui, Doroud:2012xw, Closset:2015rna, Hori:2014tda}.\\

\noindent
\ul{\large Vector multiplet $\Sigma$, Ghost sector}\\

The relevant terms in the vector multiplet and the ghost sector are 
\bal
 \mc{Q}V_\Sigma &= \frac12 \Bigl( F_{\hat{1}\hat{2}} - \frac{\eta}{r} \Bigl)^2 + \frac12\Bigl( D + \frac{\sigma}{r} \Bigl)^2 + \frac12(\nabla\sigma)^2 + \frac12(\nabla\eta)^2 + \frac i2 \lambda^\dag\gamma^m\nabla_m\lambda\ , \label{eq:qvsigma}
 \\[5pt]
 \mc{L}_{\rm ghost} &= - \ol{c} \nabla^2 c - \frac{1}{2\xi}(\nabla^mA_m)^2\ . \label{eq:lghost}
\eal
The expansion of the fields around the saddle-point is given as 
\bal
 A_m = A_{0,m} + \frac{1}{\sqrt{t}}\what{A}_m, \hs{10pt} D = - \frac{\sigma_0}{r} + \frac{1}{\sqrt{t}}\what{D} \hs{10pt} \sigma = \sigma_0 + \frac{1}{\sqrt{t}}\what{\sigma}, \hs{10pt} \eta = \eta_0 + \frac{1}{\sqrt{t}}\what{\eta}, \hs{10pt} \lambda = \frac{1}{\sqrt{t}}\what{\lambda} \ ,
\label{eq:expandghostvector} 
\eal
where $\sigma_0$ and $\eta_0$ parametrize the moduli space on the Coulomb branch. Hereafter we will omit the hat for fluctuations for a notational simplicity. 
$A_{0,m}$ in (\ref{eq:expandghostvector}) is a solution to the saddle-point equation (\ref{eq:saddle1}), $F_{\hat{1}\hat{2}} = \frac{\eta_0}{r} = \frac{m}{2r^2}$ with $m \in \mbb{Z}$, 
and we choose the Wu-Yang monopole solution \cite{Dirac:1931kp, Wu:1975es, Wu:1976ge}: 
\[
A_{0,\theta} =0\ , \ \ \  
A_{0,\varphi} = 
\left\{
\begin{array}{ll} 
\eta_0(1 - \cos\theta ) & \mbox{if}\ \ \ 0 \le \theta \le \theta_0 \\
-\eta_0(1 + \cos\theta ) & \mbox{if}\ \ \ \theta_0 \le \theta \le \pi
\end{array}
\right. 
\ , 
\label{eq:monopole}
\] 
where $0 < \theta_0 < \pi$. The north-pole and the south-pole patches that respectively cover the regions, 
$U_N = \{ (\theta, \varphi) | 0 \le \theta \le \theta_0,\ 0 \le \varphi \le 2\pi \}$ and $U_S = \{ (\theta, \varphi) | \theta_0 \le \theta \le \pi,\ 0 \le \varphi \le 2\pi \}$, 
are stitched together at $\theta = \theta_0$. 
Note that from (\ref{eq:lghost}), in the large-$t$ limit, the gauge field should satisfy the Lorentz-gauge condition, $\nabla^m A_m=0$, 
and the Wu-Yang monopole solution (\ref{eq:monopole}) indeed satisfies the condition.

We focus on the bosonic part of the vector multiplet. 
The integral over the auxiliary field $D$ will be treated when we discuss the S\"uckelberg chiral multiplet. 
The rest of the bosonic part can be represented as a matrix form: 
\begin{align}
 \mc{Q}V_\Sigma^b &= \frac12 \Bigl( F_{\hat{1}\hat{2}} - \frac{\eta}{r} \Bigl)^2 + \frac12( \nabla_m\sigma )^2 + \frac12( \nabla_m\eta )^2 - \frac{1}{2\xi}(\nabla^mA_m)^2  \nn\\
 							&=: \left( \begin{array}{cccc}
										A_+ & A_- & \eta & \sigma 
									\end{array} \right) \mc{O}_{\rm \Sigma}^{b} 
								\left( \begin{array}{c} 
										A_+ \\
										A_- \\
										\eta \\
										\sigma
									\end{array} \right)\ , 
\end{align}
where we choose $\xi=-1$; 
$A_\pm:=(A_1 \mp iA_2)/2$; 
$F_{mn} = \nabla_mA_n - \nabla_nA_m = \del_mA_n - \del_nA_m$.
As before, let us do the case study in terms of the value of $j$. 
\\

\begin{enumerate}

\item{$1 \le j$}

In this case, the operator $ \mc{O}_{\rm \Sigma}^{b}$ is a $4\times 4$ matrix: 
\begin{align}
 \mc{O}_{\rm \Sigma}^{b} &= \left( \begin{array}{cccc}
										- 4\nabla_+\nabla_- & 0 & - \frac ir \nabla_+ & 0 \\
										0 & - 4\nabla_-\nabla_+ & +\frac ir \nabla_- & 0 \\
										- \frac ir \nabla_- & +\frac ir \nabla_+ & - \frac12\nabla^2 - \frac1{2r^2} & 0 \\
										0 & 0 & 0 & - \frac12\nabla^2 \\
									\end{array} \right). 
\end{align}
The vector-field part is diagonal due to the gauge-fixing term. 
Since the scalar fields do not affect the background monopole, we have 
\begin{align}
{\rm det}~\mc{O}_{\rm \Sigma}^{b} &= \frac12 \Bigl[ j(j+1) \Bigl]^2 \cdot \Bigl[ j(j+1) \Bigl]^2\ .
\end{align}

\item{$j = 0$}

In this case, we focus only on the scalar harmonics:
\begin{align}
 \mc{O}_{\rm \Sigma}^{b} 
						&= \left( \begin{array}{cc}
										\frac{1}{2} j (j+1) + \frac1{2} & 0 \\
										0 & \frac{1}{2} j (j+1) \\
									\end{array} \right)\ . 
\end{align}
Therefore, the eigenvalues are 
$ \frac12$ and $0$. The zero-mode for $\sigma$, $\sigma_0$, is important since it corresponds to the Coulomb branch moduli, 
but an integral over $\sigma_0$ will be localized as we will see in (\ref{eq:deltasigma0}). 
\end{enumerate}

Since the ghost field is a Grassmann-odd scalar, the eigenvalues are equivalent to those of the scalar case:  
\bal
 {\rm Det}' ~\mc{O}_\Sigma^{c, \ol{c}} = \prod_{j=1} \Bigl( j(j+1)  \Bigl)^{2j+1}\ ,
\eal
where the prime means to remove the zero-mode. 
In fact, the zero-mode in the ghost sector can be independently dealt with \cite{Pestun:2007rz, Ashok:2013pya}.

Concerning the $1$-loop determinant for the fermion part, 
we can carry out Gaussian integrals using the spherical harmonics, 
which yields 
\bal
 {\rm Det} \ \mc{O}_\psi &= \prod_{j=\frac12} \biggl[ \Bigl( j + \frac12 \Bigl)^2 \biggl]^{2j+1}\ . 
\eal

As a result, the $1$-loop determinant for the vector multiplet is
\bal
Z_\Sigma = {\rm constant} \ .
\eal

\noindent
\ul{\large S\"uckelberg chiral multiplet $\Gamma$}\\

Let us evaluate the $1$-loop determinant for the S\"uckelberg-type chiral multiplet. 
Although we do not construct deformation terms for the chiral multiplet $\Gamma$, 
the Lagrangian density of this multiplets is already quadratic. 
Therefore, we can perform the Gaussian integrals at the saddle-point \cite{Ashok:2013pya, Harvey:2014nha}.
The Lagrangian density for the S\"uckelberg-type chiral multiplet $\Gamma$ at the saddle-point is 
\bal
 \mc{L}_\Gamma |_{\rm saddle} &= \frac1{g^2} \Bigl[ \frac12 (\nabla_m r^3)^2 - i \ol{\wtilde{\chi}} \gamma^m \nabla_m\wtilde{\chi} \Bigl] + \frac{g^2}2(D_m\gamma^4)^2 
 +g^2 |G_{\Gamma}|^2 + \frac{g^2}2(\sigma_0^2 + \eta_0^2) + i\sqrt{2}r^3_0 D_0\ .
 \label{eq:lgamma}
\eal

One can evaluate the $1$-loop determinant for the S\"uckelberg-type chiral multiplet as we did for other multiplets, 
but there exist two parts to watch out for: 
The first part that we need to handle with care is the one that couples the non-compact scalar zero mode $r^3_0$ with $D_0$, a ``zero mode'' or a constant mode of $D$.  
The integrals associated with this coupling term can be evaluated together with a term appearing in the vector multiplet (\ref{eq:qvsigma}): 
\bal
 \int \mc{D}D\ dr^3_0 ~e^{- t (D + \frac{\sigma}{r})^2 - i\sqrt{2}(r^3_0 - t_1) D_0 }. 
 \label{eq:dddr3}
\eal
The $(D+\frac{\sigma}{r})^2$-term can be decomposed in terms of the angular momentum $l$:
\bal
 - t \left( D + \frac{\sigma}{r} \right)^2 
 	&= - t \left( D_0 + \frac{\sigma_0}{r} \right)^2 
	- \left( D + \frac{\sigma}{r} \right)^2\biggl{|}_{l=1} 
	- \left( D + \frac{\sigma}{r} \right)^2\biggl{|}_{l=2} 
	+ \cdots \ ,
\label{eq:decomposition}
\eal
where the $t$-dependence originates with the expansion (\ref{eq:expandghostvector}). 
The first term in (\ref{eq:decomposition}) is zero due to the saddle-point condition for the vector multiplet, 
i.e. 
\bal
D_0 = - \frac{\sigma_0}{r}\ , 
\label{eq:d0}
\eal
which means that the integrations over $D_0$ and $\sigma_0$ are done so as to satisfy (\ref{eq:d0}). 
The integrations over $D$ with $l \ge 1$ can be trivially done. 
Through the manipulations above, the integral (\ref{eq:dddr3}) can be evaluated as follows:  
\bal
 \int \mc{D}D\ dr^3_0 ~e^{- t (D + \frac{\sigma}{r})^2 - i\sqrt{2}(r^3_0 - t_1) D_0] } 
 \propto \int dr^3_0 ~e^{ - i\sqrt{2}(r^3_0 - t_1) (-\frac{\sigma_0}{r}) } 
 \propto \delta(\sigma_0) \ . 
 \label{eq:deltasigma0}
\eal
Therefore, the Coulomb branch moduli is fixed due to the appearance of the delta function as claimed.

The second part that we need to take care is the compact-scalar term $\frac{g^2}2(D_m\gamma^4)^2$, 
in which we need to find the classical solution of $\gamma^4$ so as to satisfy 
$(i)$ the gauge-invariance of $D_m\gamma^4$, 
$(ii)$ the equation of motion of $\gamma^4$,  
and $(iii)$ the finiteness of the integral of $(D_m\gamma^4)^2$ over $S^2$. 
The condition $(i)$ yields, at $\theta = \theta_0$,  
\bal
D_m\gamma^4 \bigl{|}_{U_N} = D_m\gamma^4 \bigl{|}_{U_S}\ ,   
\label{eq:gaugeinvariance}
\eal   
where the two patches covering the regions, $U_N = \{ (\theta, \varphi) | 0 \le \theta \le \theta_0,\ 0 \le \varphi \le 2\pi \}$ and $U_S = \{ (\theta, \varphi) | \theta_0 \le \theta \le \pi,\ 0 \le \varphi \le 2\pi \}$, 
respectively are stitched together at $\theta = \theta_0$.   
This is because the gauge field evaluated in the north-pole patch can be transformed into the one evaluated in the south-pole patch 
via a gauge transformation at $\theta = \theta_0$, and the quantity $D_m\gamma^4$ is gauge-invariant.  

In order to satisfy all the conditions $(i)$--$(iii)$, for a given non-zero $g$, we find that the following restriction should be imposed on $\eta_0$ (see appendix \ref{sec:compactscalar} for the detail): 
\[
\eta_0 = \frac{m}{2r} = 0\ , 
\label{eq:eta0}
\] 
which means that the integer $m$, the first Chern class, is selected to be zero. 
To circumvent the vanishing $\eta_0$, one has to set $g=0$, 
through which one can extract non-zero instanton contributions to the sphere partition function. 
With this understanding, in (\ref{eq:lgamma}) we set the compact-scalar term, $\frac{g^2}2(D_m\gamma^4)^2$, to zero, 
and concerning the auxiliary-field term, $g^2|G_{\Gamma}|^2$, as well as the $\mathcal{O}(g^{-2})$ terms, we first implement Gaussian integrations for a finite $g$ and then take the small-$g$ limit in the end.   
In this manipulation, unimportant overall factors appear and we absorb them into the definition of the integral measure, $\mathcal{D}\gamma^4$, as follows: 
\bal
\lim_{g \to 0} \frac{ {\rm Det}\left(-\frac{i}{g^2}\gamma^m\nabla_m\right)  }{ \sqrt{  {\rm Det}'\left(-\frac{1}{g^2}\nabla^2\right)}} \frac{1}{g^2} \prod^{\infty}_{l=1} \left( \frac{1}{g^2} \right)
\int \mathcal{D}\gamma^4 := 1\ ,
\label{eq:defmeasures} 
\eal  
where the prime means to remove the zero-mode.

As a result, the $1$-loop determinant for the S\"uckelberg-type chiral multiplet evaluated on the Coulomb branch 
becomes 
\bal
 Z_{\Gamma}
					\propto \delta (\sigma_0)\ ,
					\label{eq:zgamma1loop}
\eal
where $g$ is set to zero to obtain non-trivial instanton contributions.

\section{Sphere partition function}
\label{sec:result}

We write down the exact sphere partition function of GLSM for the KK$5$-brane with a single $U(1)$ gauge group on the Coulomb branch and with $g$ set to zero\footnote{For the multi-centered case, see the appendix \ref{sec:multi}.}:
\bal
Z_{\rm GLSM}^{S^2}&= \int \mc{D}\Sigma\ \mc{D}\Phi\ \mc{D}\Psi\ \mc{D}\Gamma\ \mc{D}\wtilde{Q}\ \mc{D}Q ~ e^{- \frac{1}{2\pi}S_{\rm KK5} - t \mc{Q}V} 
		= \sum_{m \in \mbb{Z}} e^{-i \sqrt{2} t_2 m }  Z_{\wtilde{Q}} Z_Q\ , \label{eq:zsingle}
\eal
where $S_{\rm KK5}$ denotes the action of the $U(1)$ GLSM for a single-centered KK$5$-brane on $S^2$ on the Coulomb branch;  
the $1/2\pi$ factor is a normalization constant of the world-sheet sigma model; 
\bal
 Z_{\wtilde{Q}} = \frac{ \Gamma\Bigl( - \frac{m + n_f}{2} - i\sigma_f \Bigl) }{ \Gamma\Bigl(1 - \frac{m + n_f}{2} + i\sigma_f \Bigl)}\ ,\ \ \ 
 Z_{Q} = \frac{ \Gamma\Bigl( \frac{m + n_f}{2} + i\sigma_f \Bigl) }{ \Gamma\Bigl(1 + \frac{m + n_f}{2} - i \sigma_f  \Bigl)}\ . 
\eal
Here $t_2$ is the $\theta$-parameter and $m$ is the first Chern class.

Let us read off quantum aspects of the Taub-NUT space from the exact partition function (\ref{eq:zsingle}) for the single-centered case.  
In particular, we interpret non-trivial contributions from the saddle-point as the world-sheet instanton effects. 
This means that we consider the partition function of GLSM as that of the world-sheet theory. 
In the context of GLSM, the IR limit means to tune the gauge coupling constant $e$ to infinity, 
and since the vector multiplet is $Q$-exact, the partition function does not depend on $e$, i.e. the partition function is invariant under the renormalization-group flow. 
From the point of view of the world-sheet theory obtained through the IR limit of GLSM, one can understand that the $i \sqrt{2} t_2 m$-term in the resultant partition function (\ref{eq:zsingle}) originates with the contribution from the NS $2$-form.

We first review the world-sheet instanton effects in the single-centered and the multi-centered KK$5$-brane geometry, i.e. the Taub-NUT space \cite{Okuyama:2005gx, Harvey:2005ab, Kimura:2018hph, Gregory:1997te}. 
In this article, we call the world-sheet instantons in the single- and multi-centered cases the disk instantons and the $2$-cycle instantons, respectively; 
the origin of the names will be clarified in due course.

\begin{table}[htbp]
\begin{minipage}{1.0\hsize}
  \begin{center}
   \begin{tabular}{c||c|c|c|c|c|c|c|c|c|c} \hline
            dim & $x^0$ & $x^1$ & $x^2$ & $x^3$ & $x^4$ & $x^5$ & $x^6$ & $x^7$ & $x^8$ & $x^9$  \\ \hline \hline
             KK5 & $\odot$ & $\odot$ & $\odot$ & $\odot$ & $\odot$ & $\odot$ & - & - & - & $\wtilde{\ast}$ \\\hline
          \end{tabular}
          \caption{The brane configuration: The directions that the brane extends are denoted by $\odot$, and the compact direction ($\wtilde{S}^1$) by $\wtilde{\ast}$.}
          \label{tab:braneconfiguration}
  \end{center}
   \end{minipage}
\end{table}

In general, the Taub-NUT space is described by the following metric\footnote{Concerning the harmonic function $H(R)$, we follow the convention in \cite{Kimura:2013fda}.}:
\bal
 ds^2_{\text{Taub-NUT}} &= H(R)dx_{\mbb{R}^3}^2 + \frac{1}{H(R)} ( d\theta + \omega_{i}dx^i )^2 \ , \hs{20pt} (i=6,7,8) \ , \label{eq:taubnut} \\
 H(R) &= \frac{1}{g^2} + \sum_a\frac{1}{\sqrt{2}R_a} \ , \ \ \ \text{with}\ \ \ 
 R_a = \sqrt{(x^6 - s^1_{a})^2 + (x^7 - s^2_{a})^2 + (x^8 - t^1_{a})^2} \ . \label{eq:r}
\eal
Here the Taub-NUT space is transverse to the KK$5$-branes; 
$dx^2_{\mbb{R}^3}$ represents a metric on the base space $\mbb{R}^3$; 
 $\theta$ is a coordinate on a compact direction (see Table \ref{tab:braneconfiguration}). 
 This Taub-NUT space is locally described as $\mbb{R}^3 \times S^1$ and is an $S^1$-fibration over the base space $\mbb{R}^3$.      
 The structure of this $S^1$-fibration can be understood from the harmonic function $H(R)$ where $R_a$ denotes the distance from the center (see (\ref{eq:r})).  
$s^1_a$, $s^2_a$ and $t^1_a$ are the moduli parameters that specify the location of the centers, and they correspond to the FI-parameters in the context of GLSM.

The harmonic function $H(R)$ has the following asymptotic behaviors at the centers and at infinity:
\bal
 \frac1{H(R\rightarrow0)} = 0, \hs{20pt} \frac1{H(R\rightarrow\infty)} = g \ ,
\eal
which means that  the $S^1$ fiber at the centers (at $R=0$) shrinks to zero size, and it asymptotically becomes $g$ at infinity. 
The metric (\ref{eq:taubnut}) may look singular at the centers, but in fact the geometry becomes $\mbb{R}^4$ at the centers by a proper coordinate transformation.    

In the case of the multi-centered Taub-NUT space, there exist $S^2$'s swept out by the circle fiber as it moves from one center to the other, 
and therefore, the existence of $2$-cycles can define the world-sheet instantons in the Taub-NUT space:   
\bal
 \phi \ : \ \Sigma = S^2 \longrightarrow S^2, \hs{20pt} \pi_2(S^2) = \mbb{Z}\ . 
\eal

At first sight, in the single-centered case, one may think that nothing drastic happens because in the context of GLSM, one obtains the single-centered case from the multi-centered case by selecting a single $U(1)$. 
However, it is important that the geometric picture drastically changes. 
The single-centered case can be obtained from the two-centered case by putting one center to infinity, 
i.e. it forms an open-cigar geometry that is topologically $D^2$. 
This means that there is no non-trivial $2$-cycle in the single-centered Taub-NUT space. 
In this regard, however, it is also known that there exist the world-sheet instanton effects in the single-centered case as in the multi-centered case \cite{Harvey:2005ab}. 
This makes it difficult to understand the world-sheet instantons in the single-centered Taub-NUT space.

The instantons for the single-centered case discussed in \cite{Harvey:2005ab} have been contemplated in \cite{Okuyama:2005gx}. 
The proposal given by \cite{Okuyama:2005gx} is to decompose the world-sheet topology as    
\bal
 \Sigma = S^2 = D^2 \cup \{ \infty \} \ . 
 \label{eq:dtimesinf}
\eal
Through this world-sheet disk, one can define instantons. 
The disk instanton introduced in \cite{Okuyama:2005gx} is given by the following map:
\bal
 \Sigma \ni D^2 := \{ z  |\ |z| < 1 \}\ \to\ nC_a := \{ \vec{r}(z) = \vec{r}_a + f( |z| )\vec{v}, \ \theta(z) = n \cdot {\rm arg}(z) \}\ ,
 \label{eq:diskinstanton}
\eal
where $C_a$ denotes a cigar that shrinks at $\vec{r}_a$; 
$\vec{v}$ is a unit vector specifying the direction of cigar; 
 $f(z)$ is a function satisfying the following boundary condition:  
\bal
 f(0) = 0\ , \ \ \  f(1) = \infty\ .
\eal
If going around the origin in the world-sheet at $z=z_0$ with $|z_0|<1$, it wraps $n$ times on the semi-infinite cigar. 
Note here that the map (\ref{eq:diskinstanton}) is well-defined for arbitrary $g$. 

In \cite{Kimura:2018hph}, based on the map (\ref{eq:diskinstanton}) for the single-centered case, 
the relation between the disk instantons and the $2$-cycle instantons was discussed in the $g\to 0$ limit: 
If taking the $g\to 0$ limit in the single-centered case, the $S^1$ fiber at infinity would shrink, and accordingly the target-space topology would change from $D^2$ to $S^2$, 
i.e. the disk instantons become the $2$-cycle instantons in the $g\to 0$ limit\footnote{In this regard, however, it may be the case that the topology change would not occur in the $g\to 0$ limit, i.e. the disk instantons stay as the disk instantons even in the $g\to 0$ limit. In order to understand the relation between the $2$-cycle instantons and disk instantons in the $g\to 0$ limit, 
it would be important to understand how the target-space geometry changes in the limit.}.

Let us contemplate the world-sheet instantons appearing in the sphere partition function for the single-centered case (\ref{eq:zsingle}).  
The instanton calculus has been done restrictively in the $g\to 0$ limit \cite{Tong:2002rq}, and it is known to break the isometry in the target space of the KK$5$-brane in the GLSM \cite{Okuyama:2005gx, Harvey:2005ab}. 
The portion $e^{-i \sqrt{2} t_{2} m }$ appearing in the sphere partition function (\ref{eq:zsingle}) for $g=0$ 
is indeed a contribution that appears as the world-sheet instantons with $t_{2}$ as the $\theta$-parameter in \cite{Okuyama:2005gx, Harvey:2005ab}.  
In the single-centered case, we identify these instantons with the disk instantons introduced in \cite{Okuyama:2005gx}.

Next, we consider the world-sheet instanton effects in a finite-$g$ region. 
Evaluating the $1$-loop determinant for the S\"uckelberg-type chiral multiplet, 
we have shown that for a finite value of $g$, the only possible instanton number should be zero due to the three conditions $(i)$--$(iii)$ discussed in section \ref{sec:determinant}. 
This means a constant map from the world-sheet to the target space, i.e. no winding exists. 
Therefore, we conclude that the world-sheet instanton does not exist for a finite value of $g$.

We make a comment on the FI-parameter part that is also a contribution from the world-sheet instanton of the GLSM \cite{Tong:2002rq, Okuyama:2005gx, Harvey:2005ab}. 
Since the scalar kinetic term in the S\"uckelberg-type multiplet is frozen at small $g$, the dynamical FI-parameters are localized 
to give the distance from the centers.  
However, this contribution disappears in the supersymmetric localization since the integrations over dynamical fields are performed explicitly.

\section{Summary}
\label{sec:metric}

In this section, we overview what we have done and discuss the outcome. 
Simply stated, for the purpose of investigating the stringy effects, we have calculated the sphere partition function of GLSM for the KK$5$-brane with a single $U(1)$ gauge group on the Coulomb branch using the supersymmetric localization; 
from the result obtained, we have discussed the world-sheet instanton effects.  

It is known that in the context of GLSM, the background geometries of the H-monopoles and the KK$5$-branes are corrected by the world-sheet instantons \cite{Tong:2002rq, Gregory:1997te}, 
which means that one can deal with the world-sheet instantons based on GLSM that allows us to conduct analyses from the view point of string.  
In particular, we have considered GLSM for the KK$5$-branes \cite{Harvey:2005ab} together with the introduction of the moduli specifying the location of the branes in the target space \cite{Okuyama:2005gx}.

Let us focus on calculations that the St\"uckelberg-type chiral multiplet is involved: 
The St\"uckelberg-type chiral multiplet carries weight in the discussion about the world-sheet instantons because it describes the $S^1$-fiber part of the Taub-NUT space. 
It is important to mention that evaluating integrals over the zero-mode of the non-compact scalar in the St\"uckelberg-type chiral multiplet 
yields a delta-function 
that selects the vacuum expectation value of the adjoint scalar field in the gauge multiplet to be zero as we have seen in (\ref{eq:deltasigma0}). 
In addition, analyzing the kinetic term of the compact scalar field in St\"uckelberg-type chiral multiplet, we have shown that the first Chern class $m$ is fixed to zero for any finite value of $g$. 
This result means that the embedding map from the world sheet onto the target space, the Taub-NUT space, only exists as a constant map, i.e. 
the world sheet does not wind around the Taub-NUT space. 
In addition, setting that $g=0$, it is possible to find out contributions from non-trivial instanton numbers to the sphere partition function (of GLSM for the KK$5$-brane with a single $U(1)$ gauge group), 
and we have concretely worked this out.

 The result of the sphere partition function of GLSM obtained in this study shows the existence of non-trivial contributions from the saddle points.        
In particular, we have discussed the physical meaning of the instanton number based on the results and interpretation in \cite{Harvey:2005ab, Okuyama:2005gx}. 
In the discussion given in \cite{Harvey:2005ab, Okuyama:2005gx}, the contributions from the world-sheet instantons have been obtained in the $g \to 0$ limit, 
by searching for the instanton configurations of GLSM. 
In fact, the contributions in the sphere partition function studied in this article are those that appear in the sphere partition function at $g=0$, 
which is consistent with the previous studies \cite{Harvey:2005ab, Okuyama:2005gx}. 
Therefore, comparing with the interpretation of the world-sheet instanton given in \cite{Okuyama:2005gx}, 
the instanton contributions in our case can be understood as follows.     
In the single-centered Taub-NUT space, the topology of the $S^1$ fiber is $D^2$. 
In this case, decomposing the world-sheet $S^2$ into $D^2 \cup \{ \infty \}$, the world-sheet disk can wind around the target space $D^2$ in a non-trivial manner, 
which is the disk instanton given in \cite{Okuyama:2005gx}.  

Although we do not go into detail, we have calculated the sphere partition function for the multi-centered case in the appendix \ref{sec:multi}, 
which shows that the non-trivial contributions from the saddle-points do exist even for a finite value of $g$. 
Clarifying the physical interpretation of finite-$g$ instanton contributions for the multi-centered case is a quite important future direction. 
As well, in order to understand the world-sheet instantons quantitatively, it would be important to clarify how the geometry of the Taub-NUT space changes in the $g\to 0$ limit. 
We make a comment on stringy corrections for the moduli space for the Taub-NUT space. 
Since the classical K\"ahler metric has been obtained in \cite{Imamura:1997ss}, in principle its stringy corrections can be read off from the sphere partition function of GLSM, $Z_{\rm GLSM}^{S^2} \sim e^{-K}$ where $K$ is the K\"ahler potential \cite{Jockers:2012dk,Gomis:2012wy, Gerchkovitz:2014gta}, 
and this is also an important future work.

\vs{20pt}

\section*{Acknowledgement}
We especially thank Tetsuji Kimura for his helpful suggestions to work on the sphere partition function in relation to the K\"ahler potential, 
and for sharing private notes in the early stage.  
As well, we are grateful to 
Kazumi Okuyama,  
Tadakatsu Sakai, 
and Shin Sasaki   
for useful comments on the draft. 
We also thank 
Masashi Hamanaka, 
Keita Nii, 
Yuji Sugimoto,
and 
Sotaro Sugishita
for fruitful discussions and encouragements. 
The work of YS was supported by Building of Consortia for the Development of Human Resources in Science and Technology,  
and by JSPS KAKENHI Grant Number 19K14705.

\appendix 

\section{Convention}
\label{sec:convention}

We set up the convention that we use in this article.

Let us use various indices in the following manner: 
\begin{align}
 \left. 
 	\begin{array}{ll}
  		m,n,\cdots = 1,2 ~~&~~ \mbox{for curved space indices} \\
		a,b, \cdots = \hat{1},\hat{2} ~~&~~ \mbox{for tangent space indices} \\
		\alpha, \beta, \cdots = +,- ~~&~~ \mbox{for spinor indices} 
  	\end{array}
 \right.
\end{align}

We introduce coordinates $(\theta, \varphi)$ on the two-dimensional sphere $S^2$. 
The metric of $S^2$ whose radius is $r$, differential operators on $S^2$ and the spin connection are given as follows:
\bal
 ds^2 = r^2( d\theta^2 + \sin^2\theta d\varphi^2 ) = g_{mn}dx^m dx^n = e^ae^a\ ;
\eal
\bal
\del_{\hat{1}} = { e_{\hat{1}} }^1\del_1 = \frac1r \del_\theta\ , \hs{20pt} \del_{\hat{2}} = { e_{\hat{2}} }^2\del_2 = \frac1{r\sin\theta} \del_\varphi\ , \hs{20pt} \omega^{\hat{1}\hat{2}} = - \cos\theta d \varphi\ . 
\eal
Also the $2$-rank anti-symmetric epsilon symbol is defined as $\varepsilon_{+-}=\varepsilon^{+-}=1$.

The $2$d Clifford algebra is described by the Pauli matrices: 
\bal
 \gamma^a := \sigma^a, ~~\gamma_3 := -i\gamma^{\hat{1}}\gamma^{\hat{2}}\ , 
 ~~ \{ \gamma^a, \gamma^b\} = 2\delta^{ab}{\bf 1}\ , 
 ~~ \gamma^{ab} := \frac12(\gamma^a\gamma^b - \gamma^b\gamma^a)\ ,
\eal
where 
\begin{align}
 \gamma^{\hat{1}} = \left( \begin{array}{cc} 0 & 1 \\ 1 & 0 \end{array} \right)\ , \hs{20pt} 
 \gamma^{\hat{2}} = \left( \begin{array}{cc} 0 & -i \\ i & 0 \end{array} \right)\ , \hs{20pt} 
 \gamma^{\hat{3}} = \left( \begin{array}{cc} 1 & 0 \\ 0 & -1 \end{array} \right)\ . 
\end{align}
In the Euclidean space, a Dirac spinor $\psi$ and a barred Dirac spinor $\ol{\psi}$ are independent. 
We use the conjugation matrix $C$ to raise and lower spinor indices. 
In this article, we define the contraction of spinor indices as follows \cite{Benini:2012ui, Benini:2013nda, Benini:2013xpa}:
\bal
 \ol{\psi} \lambda := \ol{\psi}^\alpha \lambda_\alpha\ , \hs{20pt}  
 \ol{\psi} \gamma^a \lambda := \ol{\psi}^\alpha {(\gamma^a)_\alpha}^\beta \lambda_\beta \ . 
\eal
The charge conjugation matrix is defined as
\bal
 C^\dag = C\ , \hs{20pt} C^2 = 1\ , \hs{20pt} C^T = -C\ , \hs{20pt} C\gamma^aC = -(\gamma^a)^T\ .
\eal
Grassmann-odd and -even spinors satisfy the following relations:
\begin{align}
\left.
	\begin{array}{cc}
 		 & \psi \lambda = \lambda \psi \\
 		\mbox{Grassmann-odd} & \psi \gamma^a \lambda = - \lambda\gamma^a \psi \\
		 & \psi \gamma^a\gamma^b \lambda = \lambda \gamma^b\gamma^a \psi 
 	\end{array} \right. \hs{20pt}
	\left. \begin{array}{cc}
		 & \psi \lambda = -\lambda \psi \\
 		\mbox{Grassmann-even} & \psi \gamma^a \lambda = + \lambda\gamma^a \psi \\
		 & \psi \gamma^a\gamma^b \lambda = - \lambda \gamma^b\gamma^a \psi
	\end{array}
\right.
\end{align}
The Fierz identity for Grassmann-even spinors is given as
\bal 
 (\epsilon^\dagger\lambda_1)\lambda_2 = \frac12 \Bigl[ \lambda_1(\epsilon^\dagger\lambda_2) + \gamma_3\lambda_1(\epsilon^\dagger\gamma_3\lambda_2) + \gamma^m\lambda_1(\ol{\epsilon}\gamma_m\lambda_2) \Bigl].
\eal

\section{Conformal Killing spinor}
\label{sec:Killing}

We use the conformal Killing spinor in order to construct supersymmetry on $S^2$, satisfying the conformal Killing spinor equations \cite{Benini:2012ui, Sugishita:2013jca, Closset:2014pda, Lu:1998nu}:
\bal
 \nabla_m \epsilon = \frac{i}{2r} \gamma_m \epsilon\ , \hs{20pt} \nabla_m \ol{\epsilon} = \frac{i}{2r} \gamma_m \ol{\epsilon}\ .
 \label{eq:ckse}
\eal 
A solution to the conformal Killing spinor equation (\ref{eq:ckse}) is \cite{Benini:2012ui}:
\begin{align}
 \epsilon = C_1 e^{-i \frac{\varphi}{2}} \left( \begin{array}{cc} \sin \frac{\theta}{2} \\ -i\cos \frac{\theta}{2} \end{array} \right) + C_2 e^{i \frac{\varphi}{2}} \left( \begin{array}{cc} \cos \frac{\theta}{2} \\ i\sin \frac{\theta}{2} \end{array} \right).
\end{align}
We choose $C_1=0$ and $C_2=1$ to perform the supersymmetric localization. 
Here we have the relation:
\bal
 \epsilon^\dag \epsilon = 1\ ,\hs{20pt} \epsilon^T C \epsilon = 0\ .
\eal
A Killing vector satisfies 
\bal
 V^m = \epsilon^\dag \gamma^m \epsilon\ , \hs{20pt} \nabla_m V^m = 0\ .
\eal

\section{Spherical harmonics}
\label{sec:harmonics}

We write down a number of formulae useful for computing the $1$-loop determinants. 
We follow the convention used in \cite{Benini:2012ui}.
Spherical harmonics are the eigenfunctions of the Laplacian on $S^2$:
\bal
 \nabla^2 Y_{jm} = - \frac{j(j+1)}{r^2} Y_{jm} \ ,
\eal
where $j \in \mbb{N}$ and $ -j \le m \le j$. 
This scalar spherical harmonics $Y^{s=0}_{jm}$ is well-known. 
However, the theory we consider has various fields with spin, 
which can be dealt with the spin spherical harmonics.

A field charged under an Abelian gauge symmetry has the covariant derivative:
\bal
 D_m \Phi = (\del_m + is_0\omega_m - iQA_m) \Phi \ ,
\eal
where $\Phi$ is an arbitrary field; $s_0$ is spin; $\omega_m$ is a spin connection; $Q$ is a gauge charge; $A_m$ is a gauge connection. 
The gauge field in this article takes the monopole configuration at the saddle-point. 
Accordingly, the spin connection and the gauge field are written in terms of $\cos\theta$, and an effective spin is defined as $s_{\rm eff} := s_0 - \frac{m}{2}$ \cite{Benini:2015noa}, 
where $m$ is a monopole charge at the saddle-point.
In the case of general gauge group, the corresponding effective spin is described in \cite{Benini:2012ui}.
When evaluating eigenvalues of differential operators, we set 
\bal
 D_\pm := \frac12 ( D_1 \mp i D_2 ) \ .
\eal
These operators map the spin $s$ to $s \pm 1$ for the spin spherical harmonics $Y^s_{jm}$.
Spherical harmonics with the effective spin have the eigenvalue of the covariant derivative \cite{Benini:2012ui, Honda:2015yha}: 
\bal
 D^2 Y^{s_{\rm eff}}_{jm} = - \frac{j(j+1) - s^2_{\rm eff} }{r^2} Y^{s_{\rm eff}}_{jm},
\eal
where $j \in \mbb{N}$, $ -j \le m \le j$ and $|s_{\rm eff}| \le j$. 
A vacuum expectation value of $\eta$ at the saddle-point is associated with a magnetic flux $\frac{1}{2\pi}\int_{S^2} F = m$. 
Therefore, this $m$ is $\mbb{Z}$-valued by the charge quantization condition.

As well we consider the Dirac operator for fermionic fields. 
Spherical harmonics with the effective spin have the eigenvalues of the covariant derivative: 
\bal
 D_+ Y^{s_{\rm eff}}_{jm} = \frac{s^+_{\rm eff} }{2r} Y^{s_{\rm eff}+1}_{jm}, \hs{20pt} D_- Y^{s_{\rm eff}}_{jm} = - \frac{s^-_{\rm eff} }{2r} Y^{s_{\rm eff}+1}_{jm}\ ,
\eal
where the eigenvalues $s^\pm_{\rm eff}$'s are 
\bal
 s^{\pm}_{\rm eff} = \sqrt{ j(j+1) - s^+_{\rm eff}( s^+_{\rm eff} \pm 1 ) }\ .
\eal
Thus using these spin spherical harmonics, we can evaluate $1$-loop determinants for all cases.

\section{Supercharges}
\label{sec:Supercharges}

The supersymmetry transformation on $S^2$ is given by
$\delta = \delta_\epsilon + \delta_{\ol{\epsilon}} = \epsilon^\alpha Q_\alpha + \ol{\epsilon}^\alpha Q^\dagger_\alpha$. 
Then susy parameters, $\epsilon$ and $\ol{\epsilon}$, satisfy the conformal Killing spinor equations as constraints on $S^2$. 
$\delta$ is a commuting operator and $\epsilon, \ol{\epsilon}$ are Grassmann-odd spinors. 
We extract Grassmann-odd susy parameters from $\delta$, and employ Grassmann-even solutions to the conformal Killing spinor equation as susy parameters. 
These obey some relations $\epsilon^\dagger\epsilon=1, ~\epsilon\epsilon = \epsilon^TC\epsilon=0$. 
Therefore, we obtain supercharges
\bal
 \mc{Q} := Q + Q^\dagger\ , \hs{20pt} Q := \epsilon^\alpha Q_\alpha, \hs{20pt} Q^\dagger := {\epsilon^c}^\alpha Q^\dagger_\alpha \ , 
\eal
where the operator $\mc{Q}$ is Grassmann-odd; $\epsilon^c$ is defined as
\bal
 \epsilon^c := C\epsilon^*\ , \hs{20pt} {\epsilon^c}^\dagger = \epsilon^TC\ .
\eal
In this way, we obtain the supersymmetry transformations with rewriting barred spinors $\ol{\psi} = C(\psi^\dag)^T$ \cite{Benini:2012ui} (also see \cite{Benini:2013yva, Benini:2016hjo}) :\\

\noindent
{\large Vector multiplet:}
\begin{align}
\begin{aligned}
 \mc{Q}A_m &= \frac i2 (\lambda^\dagger\gamma_m\epsilon + \epsilon^\dagger\gamma_m\lambda)\ , \\
 \mc{Q}\sigma &= -\frac12 ( \lambda^\dagger\epsilon + \epsilon^\dagger\lambda )\ , \\
 \mc{Q}\eta &= \frac i2 (\lambda^\dagger\gamma_3\epsilon + \epsilon^\dagger\gamma_3\lambda)\ , \\
 \mc{Q}\lambda &= i\gamma_3\epsilon \Bigl( F_{\hat{1}\hat{2}} - \frac{\eta}{r} \Bigl) - \epsilon \Bigl( D + \frac{\sigma}{r} \Bigl) + i\gamma^m\epsilon\nabla_m\sigma - \gamma_3\gamma^m\epsilon\nabla_m\eta\ , \\
 \mc{Q}\lambda^\dagger &= -i\epsilon^\dagger\gamma_3 \Bigl( F_{\hat{1}\hat{2}} - \frac{\eta}{r} \Bigl) + \epsilon^\dagger \Bigl( D + \frac{\sigma}{r} \Bigl) + i\epsilon^\dagger\gamma^m\nabla_m\sigma - \epsilon^\dagger\gamma_3\gamma^m\nabla_m\eta\ , \\
 \mc{Q}D &= \frac i2( \epsilon^\dagger\gamma^m\nabla_m\lambda - \nabla_m\lambda^\dagger\gamma^m\epsilon ) + \frac{1}{2r}( \epsilon^\dagger\lambda + \lambda^\dagger\epsilon )\ .
\end{aligned}
\end{align}

\noindent
{\large Chiral multiplet:}
\begin{align}
\begin{aligned}
 \mc{Q}\phi &= - \epsilon^\dagger\psi\ , \\
 \mc{Q}\phi^\dagger &= \psi^\dagger\epsilon , \\
  \mc{Q}\psi &= \Bigl( i\gamma^m D_m\phi + i\sigma\phi + \gamma_3\eta\phi - \frac{q}{2r}\phi \Bigl)\epsilon + \epsilon^cF\ , \\
 \mc{Q}\psi^\dagger &= \epsilon^\dagger\Bigl( -i\gamma^m D_m\phi^\dagger + i\sigma\phi^\dagger + \gamma_3\eta\phi^\dagger - \frac{q}{2r}\phi^\dagger \Bigl) - {\epsilon^c}^\dagger F^\dagger\ , \\
 \mc{Q}F &= {\epsilon^c}^\dagger \Bigl( i\gamma^m D_m\psi - i\sigma\psi + \gamma_3\eta\psi + \frac{q}{2r}\psi - i\lambda\phi \Bigl)\ ,\\
 \mc{Q}F^\dagger &= \Bigl( -iD_m\psi^\dagger\gamma^m  - i\sigma\psi^\dagger + \psi^\dagger\gamma_3\eta + \frac{q}{2r}\psi^\dagger + i\lambda^\dagger\phi^\dagger \Bigl) \epsilon^c\ .
\end{aligned}
\end{align}

\section{Evaluation of the compact-scalar term}
\label{sec:compactscalar}

We show the detailed treatment for the compact-scalar term discussed in section \ref{sec:determinant}. 
Hereafter, we will focus on the single $U(1)$ case.  
The kinetic term for the compact scalar at the saddle-point is 
\bal
D_m \gamma^4 = \nabla_m \gamma^4 + \sqrt{2} A_{0,m}\ . 
\label{eq:dmgamma}
\eal 
This is invariant under the gauge transformation:
\bal
\gamma^4 \to \gamma^4 - \sqrt{2} \lambda\ , \ \ \ 
A_{0,m} \to A_{0,m} + \nabla_m \lambda\ , 
\eal
where $\lambda$ is an arbitrary function. 
We have chosen the gauge field at the saddle point to be the Wu-Yang monopole (\ref{eq:monopole}). 

In order to find the proper classical value of $\gamma^4$, we impose the three conditions: 
$(i)$ the gauge-invariance of $D_m\gamma^4$, 
$(ii)$ the equation of motion of $\gamma^4$,  
and $(iii)$ the finiteness of the integral of $(D_m\gamma^4)^2$ over $S^2$.  
 
The condition $(i)$ yields
\bal
D_m\gamma^4 \bigl{|}_{U_N} = D_m\gamma^4 \bigl{|}_{U_S} \ \ \ \text{at} \ \ \ \theta=\theta_0\ ,   
\label{eq:gaugeinvariance2}
\eal   
where the two patches covering the regions, $U_N = \{ (\theta, \varphi) | 0 \le \theta \le \theta_0,\ 0 \le \varphi \le 2\pi \}$ and $U_S = \{ (\theta, \varphi) | \theta_0 \le \theta \le \pi,\ 0 \le \varphi \le 2\pi \}$, 
respectively are stitched together at $\theta = \theta_0$, and $0 < \theta_0 <\pi$. 
From (\ref{eq:gaugeinvariance2}), we have in the north-pole patch, 
\begin{align}
D_{\theta}\gamma^4 &= \partial_{\theta}\gamma^4 =g^{N}(\theta,\varphi)\ , \label{eq:north1}\\
D_{\varphi}\gamma^4 &= \partial_{\varphi}\gamma^4 + \eta_0 \sqrt{2} (1-\cos \theta) =f^{N}(\theta,\varphi)\ , \label{eq:north2}
\end{align}
and in the south-pole patch, 
\begin{align}
D_{\theta}\gamma^4 &= \partial_{\theta}\gamma^4 =g^{S}(\theta,\varphi)\ , \label{eq:south1}\\
D_{\varphi}\gamma^4 &= \partial_{\varphi}\gamma^4 - \eta_0 \sqrt{2} (1+\cos \theta) =f^{S}(\theta,\varphi)\ , \label{eq:south2}
\end{align}
where $g^N$, $f^N$, $g^S$ and $f^S$ are arbitrary functions satisfying the boundary conditions:
\bal
g^N(\theta_0,\varphi) = g^S(\theta_0,\varphi)\ , \ \ \ 
f^N(\theta_0,\varphi) = f^S(\theta_0,\varphi)\ . 
\label{eq:boundaryconditions}
\eal 

The condition $(ii)$ yields 
\bal
\left( 
\frac{1}{\sin \theta} 
\partial_{\theta} (\sin \theta \partial_{\theta}) 
+ \frac{1}{\sin^2 \theta} \partial^2_{\varphi}
\right)\gamma^4 = 0\ ,
\label{eq:eom}
\eal
where we have used the fact that $\nabla^mA_{0,m}=0$.

We focus on the region $U^N$. 
From (\ref{eq:north1}) and (\ref{eq:north2}), we find 
\bal
\partial_{\varphi} g^N = \partial_{\theta}f^N - \eta_0 \sqrt{2}\sin \theta\ . 
\label{eq:north3}
\eal 
Inserting (\ref{eq:north1}) and (\ref{eq:north2}) into the equation of motion (\ref{eq:eom}), we obtain 
\bal
\cos\theta g^N  + \sin\theta \del_\theta g^N + \frac{1}{\sin\theta} \del_\varphi f^N = 0\ .
\label{eq:north4}
\eal
Acting $\partial_{\varphi}$ on (\ref{eq:north4}) and using (\ref{eq:north3}), we have 
\bal
\nabla^2_{S^2}
f^N = 2\sqrt{2}\eta_0 \cos\theta \ , \ \ \ 
\text{with}\ \ \ 
\nabla^2_{S^2} := \frac{1}{\sin\theta}\del_\theta ( \sin\theta \del_\theta ) + \frac{1}{\sin^2\theta} \del_\varphi^2\ . 
\label{eq:north5}
\eal
Once we find the solution to (\ref{eq:north5}) satisfying the condition $(iii)$, 
we can determine $g^N$ so as to satisfy (\ref{eq:north3}), (\ref{eq:north4}) and the condition $(iii)$. 

Since $f^N$ is periodic in $\varphi$ with the period $2\pi$, we expand $f^N$ in terms of the basis $\{ e^{ik \varphi} \}$ with integer $k$:
\bal
f^N (\theta,\varphi) = \sum^{\infty}_{k=-\infty} c_k(\theta) e^{ik \varphi}\ . 
\label{eq:expansion}
\eal 
Inserting (\ref{eq:expansion}) into (\ref{eq:north5}), we obtain 
\bal
\sum^{\infty}_{k=-\infty} 
\left(
\cos\theta \del_\theta c_k + \sin\theta \del^2_\theta c_k - \frac{k^2}{\sin\theta} c_k - 2\sqrt{2}\eta_0 \cos\theta\sin\theta \ \delta_{k,0} 
\right) 
e^{ik\varphi} = 0\ ,
\label{eq:north6}
\eal 
which means for each $k$, 
\bal
\cos\theta \del_\theta c_k + \sin\theta \del^2_\theta c_k - \frac{k^2}{\sin\theta} c_k - 2\sqrt{2}\eta_0 \cos\theta\sin\theta \ \delta_{k,0} = 0\ .
\label{eq:north7}
\eal

If $k=0$, the generic solution to (\ref{eq:north7}) is 
\bal
c_0 (\theta) 
= a_1 - \sqrt{2}\eta_0 \cos \theta 
+ \left(
a_2 - \frac{\eta_0}{\sqrt{2}}
\right)\log \left[ 
\tan \frac{\theta}{2}
\right]\ ,
\label{eq:solk0}
\eal 
where $a_1$ and $a_2$ are arbitrary constants. We set $a_2= \eta_0/\sqrt{2}$ since $\log \left[ 
\tan \frac{\theta}{2}
\right]$ diverges at $\theta = 0$. 
We choose $a_1$ so as to satisfy the condition $(iii)$, i.e. $c_0=\sqrt{2}\eta_0(1-\cos \theta)$.  
Similarly for the south-pole patch, we find $b_0 = -\sqrt{2}\eta_0(1+\cos \theta)$ where $f^S=\sum_k b_k e^{ik\varphi}$, 
which, however, breaks the condition $(i)$ for arbitrary $\eta_0$.  
The only solution satisfying the three conditions $(i)$--$(iii)$ is 
\bal
c_0 = 0\ , \ \ \ \text{with}\ \ \ \eta_0 = 0\ .
\label{eq:solk02}
\eal  

If $k\ne 0$, setting $x=\cos \theta$, (\ref{eq:north7}) becomes 
\bal 
\frac{d}{dx} \left( (1-x^2) \frac{d}{dx} c_k \right) - \frac{k^2}{1-x^2}c_k = 0\ . 
\label{eq:north7kne0}
\eal 
The differential equation (\ref{eq:north7kne0}) is a spacial case of the general Legendre equation, 
and the non-zero solution exists if $k=0$, which contradicts the assumption. 
Therefore, for $k\ne 0$, there is no solution. 

From the discussion above, the solution exists only if $\eta_0=0$:  
\bal 
f^N = f^S = 0\ ,  
\label{eq:solf}
\eal
where we have applied the similar argument for $f^S$ so as to satisfy the boundary condition (\ref{eq:boundaryconditions}). 
From (\ref{eq:north3}) and (\ref{eq:solf}), we find $g^N=g^N(\theta)$. Similarly we also find $g^S=g^S(\theta)$. 
Furthermore, from (\ref{eq:north4}), we obtain 
\bal
g^N = \frac{a_3}{\sin \theta}\ , 
\label{eq:solgn}
\eal
where $a_3$ is an arbitrary constant. The solution (\ref{eq:solgn}) satisfies the condition $(iii)$ only if $a_3=0$.  
Concerning $g^S$, we can follow the similar argument, which yields $g^S=0$. 

As a result, the solution satisfying the three conditions $(i)$--$(iii)$ exists only if $\eta_0=0$, which is 
\bal
f^N = f^S= 0\ , \ \ \  
g^N=g^S=0\ . 
\label{eq:solutionfg}
\eal

\section{Sphere partition function for the multi-centered case}
\label{sec:multi}

The purpose of this article is to evaluate the sphere partition function of the GLSM for the single-centered KK$5$-brane, through the use of supersymmetric localization, 
in particular for the finite-$g$ region, and as a result, we have confirmed that no non-trivial contribution exists unless we set $g=0$ in the single-centered case. 
In this appendix, we consider the multi-centered case, and sketch how to calculate the sphere partition function of the GLSM for the multi-centered KK$5$-branes. 
The procedure is almost parallel to that of the single-centered case, and let us enumerate only distinct points.  

To begin, the Lagrangian density for the St\"uckelberg-type chiral multiplet $\Gamma$ becomes (see (\ref{eq:lagrangian}))
\bal
\mc{L}_\Gamma |_{\text{saddle}} &= \frac1{g^2} \Bigl[ \frac12 (\nabla_m r^3)^2 - i \ol{\wtilde{\chi}} \gamma^m \nabla_m\wtilde{\chi} \Bigl] + g^2 \Bigl[ \frac12(D_m\gamma^4)^2 + |G_\Gamma|^2 \Bigl] \nn\\
 					&\hs{50pt}+ \frac{g^2}2 \sum_{a,b=1}^k ( \sigma_a\sigma_b + \eta_a\eta_b ) + i\sqrt{2}r^3 \sum_{a=1}^k D_a + i \sum_{a=1}^k \Bigl[ \ol{\wtilde{\chi}}\gamma_3\lambda_a + \ol{\lambda}_a\gamma_3\wtilde{\chi} \Bigl]\ . 
\eal
Accordingly, at the saddle point, (\ref{eq:deltasigma0}) is replaced by 
\bal
 \int dr^3_0 ~e^{ i2\sqrt{2}r^3_0 \sum_a (r\sigma_{0,a}) - i2\sqrt{2} \sum_a t_{1,a}(r\sigma_{0,a}) } 
 \propto e^{ - i 2\sqrt{2} \sum_a t_{1,a}(r\sigma_{0,a}) } \delta(\sigma_{0,1} + \cdots + \sigma_{0,k})\ . 
\eal

Next, evaluating the $1$-loop determinant for the St\"uckelberg-type chiral multiplet, 
we impose that the classical solution of the compact scalar $\gamma^4$ satisfies 
$(i)$ the gauge-invariance of  
\bal
D_m \gamma^4 = \nabla_m \gamma^4 + \sqrt{2} \sum_a A_{0,m,a}\ , 
\eal 
$(ii)$ the equation of motion of $\gamma^4$, and $(iii)$ the finiteness of the integral of $(D_m \gamma^4)^2$ over $S^2$ ,
which yields the conditions, $D_m \gamma^4 = 0$, and 
\bal
 \sum_{a=1}^k \frac{m_a}{2r} = \sum_{a=1}^k \eta_{0,a} = 0\ ,
 \label{eq:constraint}
\eal
for a finite value of $g$. 
Note here that each $m_a$ or $\eta_{0,a}$ can have non-zero value as opposed to the single-centered case. 
The one-loop determinant for the St\"uckelberg-type chiral multiplet evaluated on the Coulomb branch then becomes   
\bal
 Z_{\Gamma} \propto e^{ - i 2\sqrt{2} \sum_a t_{1,a}(r\sigma_{0,a}) } \delta(\sigma_{0,1} + \cdots + \sigma_{0,k})\ . 
\eal

As a result, the sphere partition function of the $U(1)^k$ GLSM for the multi-centered KK$5$-branes can be calculated as 
\bal
Z_{ {\rm GLSM}_k }^{S^2}
		&= \prod_{a=1}^k \sum_{m_a \in \mbb{Z}} \delta_{m_1+\cdots+m_k ,0} \int d(r\sigma_{0,a}) \ e^{ -i 2\sqrt{2} t_{1,a} (r\sigma_{0,a}) - i \sqrt{2} t_{2,a} m_a } \delta(\sigma_{0,1} + \cdots + \sigma_{0,k})Z_{\wtilde{Q}_a} Z_{Q_a} \ , 
		\label{eq:zmulti}
\eal
where 
\bal
 Z_{\wtilde{Q}_a} = \frac{ \Gamma\Bigl( - \frac{m_a + n_f}{2} - i\sigma_{0,a} - i\sigma_f \Bigl) }{ \Gamma\Bigl(1 - \frac{m_a + n_f}{2} + i\sigma_{0,a} + i\sigma_f \Bigl)}\ , \ \ \  
 Z_{Q_a} = \frac{ \Gamma\Bigl( \frac{m_a + n_f}{2} + i\sigma_{0,a} + i\sigma_f \Bigl) }{ \Gamma\Bigl(1 + \frac{m_a + n_f}{2} - i\sigma_{0,a} - i \sigma_f  \Bigl)}\ . 
\eal
Note that the expression of the exact partition function (\ref{eq:zmulti}) holds for a finite value of $g$, in which the $g$-dependence disappears.  
However, if we set $g=0$ from the beginning, the restriction (\ref{eq:constraint}) will never appear, and there is a discontinuity at $g=0$ as we have observed in the single-centered case.

\newpage
\addcontentsline{toc}{section}{References}
\bibliography{}

\end{document}